\documentclass[preprint,preprintnumbers,aps,pra,amsmath,amssymb]{revtex4-1}

\usepackage{graphicx}
\usepackage{amsmath}
\usepackage{bm}
\usepackage{color}

\newcommand*{\ket}[1]{\left|#1\right>}
\newcommand*{\bra}[1]{\left<#1\right|}

\newcommand*{\bok}[3]{\left<#1\left|#2\right|#3\right>}
\newcommand*{\kb}[2]{\ket{#1}\bra{#2}}

\begin{document}

\title{Characterizing Destructive Quantum Interference in Electron Transport}

\author{Panu Sam-ang}
\author{Matthew G. Reuter}
\email{matthew.reuter@stonybrook.edu}
\affiliation{Department of Applied Mathematics \& Statistics and Institute for Advanced Computational Science, Stony Brook University, Stony Brook, New York 11794, United States}
\date{\today}

\begin{abstract}
Destructive quantum interference in electron transport through molecules provides an unconventional route for suppressing electric current. In this work we introduce ``interference vectors'' for each interference and use them to characterize the interference. An interference vector may be an orbital of the bare molecule, in which case the interference is very sensitive to perturbation. In contrast, an interference vector may be a combination of multiple molecular orbitals, leading to more robust interference that is likelier to be experimentally observable. Our characterization scheme quantifies these two possibilities through the degree of rotation and also assigns an order to each interference that describes the shape of the Landauer-B\"uttiker transmission function around the interference. Several examples are then presented, showcasing the generality of our theory and characterization scheme, which is not limited to specific classes of molecules or particular molecule-electrode coupling patterns.
\end{abstract}

\maketitle

\section{Introduction}
\label{sec:introduction}
Molecules have been suggested as components of electrical circuits in the ongoing drive for device miniaturization \cite{aviram-277-1974, bk:cuevas-2010, bergfield-6412-2015}. To this end, both experimental and theoretical studies have investigated a molecule's ability to conduct electric current when sandwiched between two electrodes to form a junction \cite{bk:cuevas-2010}. In addition to such applied interest, these molecular junctions have also proven to be fundamentally valuable for investigating the mechanical strength of chemical bonds \cite{aradhya-7522-2014} and single-molecule chemical reactions \cite{li-16159-2016}.

Owing to the molecules' nanometer-scale dimensions, quantum mechanical effects are inherent in transporting electrons across these junctions. Molecular orbitals are broadened into finite-lifetime resonances when the molecule is connected to the electrodes, and the alignment of these resonances relative to the junction's Fermi energy strongly correlates with the junction's conductivity. Destructive quantum interference (DQI) is one notable exception to this principle. Some molecules (such as benzene) have multiple paths \cite{gorczak-4196-2015, solomon-621-2015} for transporting electrons across the junction, and these paths may destructively interfere with each other to suppress or even block current \cite{cardamone-2422-2006, solomon-054701-2008, hansen-194704-2009, reuter-181103-2014}. From an applications perspective, DQI may result in good molecular insulators \cite{bergfield-6412-2015}.

Since the first theoretical predictions of DQI \cite{cardamone-2422-2006, solomon-054701-2008} and subsequent experimental validation \cite{kiguchi-22254-2010, fracasso-9556-2011, taniguchi-11426-2011, guedon-305-2012, aradhya-1643-2012, arroyo-3152-2013}, numerous studies have investigated the types of molecules that exhibit DQI. Conjugated hydrocarbons are commonly employed, with DQI present in cyclic molecules, cross-conjugated molecules, and molecules with pendant groups (see \cite{reuter-181103-2014} and references therein). The common theme is that DQI primarily depends on the molecule's electronic structure and where the electrodes contact the molecule \cite{reuter-181103-2014, liu-173-2017}. For example, a benzene molecule produces DQI when connected to the electrodes in the \textit{meta} or \textit{ortho} configurations, but not \textit{para} \cite{hansen-194704-2009}.

With this observation, many guidelines have been developed for predicting molecules and molecule-electrode configurations that exhibit DQI \cite{morikawa-554-2005, pickup-198-2008, yoshizawa-9406-2008, fowler-044104-2009, fowler-244110-2009, yoshizawa-1612-2012, tsuji-224311-2014, stuyver-26390-2015}, including some graphical approaches \cite{markussen-4260-2010, markussen-14311-2011, nozaki-13951-2013}. These guidelines build physical intuition by relating DQI to either the real-space paths through the molecule (an atomic orbital-like approach) or the isolated molecule's orbitals \cite{zhao-092308-2016}. Regardless, they tend to focus only on the existence of DQI, and are most applicable to alternant hydrocarbons \cite{xia-2941-2014}, where H\"uckel or tight-binding representations for the molecule with simple molecule-electrode connections are common.

In this work, we go beyond predicting only the existence of DQI in a molecular junction and develop a broadly-applicable characterization scheme for DQI. Of primary interest is the ability to predict and classify the ``robustness'' of DQI; that is, the likelihood that the DQI will be experimentally observable. Essentially, DQI produces roots in the Landauer-B\"uttiker transmission function (\textit{vide infra}), and we recently derived an eigenvalue problem for finding these roots \cite{reuter-181103-2014}. Our main contribution here is an analysis of the corresponding eigenvectors, which we term ``interference vectors''. These interference vectors possess geometric properties that predict the line shape of the transmission function around DQI, thereby allowing us to characterize DQI.

We develop and showcase our analysis of interference vectors through several examples of increasing complexity. Our key findings include a relationship between bound states in a molecular junction and DQI in the same molecule if it were wired to the electrodes in a different configuration, the prediction of so-called supernodes \cite{bergfield-5314-2010} in oligomeric molecules, and the importance of ``coherence'' in the molecule-electrode coupling when nontrivial configurations are employed \cite{tsuji-224311-2014, hansen-6295-2016}. We also apply our analysis to DQI occurring at complex energies, which do not appear to present fundamentally new chemical insights. Our analysis is widely applicable because it builds upon a general theory of DQI \cite{reuter-181103-2014} that is not limited to conjugated hydrocarbons or simple molecular models. We thus put DQI on similar theoretical footing as a resonance analysis for locating highly-conductive molecular junctions; both analyses now possess an eigenvalue problem with physically-meaningful eigenvectors. We hope this will lead to new chemical and physical intuition for predicting, understanding, and exploiting DQI in electron transport processes.

The layout of this paper is as follows. Section \ref{sec:example} first overviews the pertinent details of Landauer-B\"uttiker theory for electron transport and then discusses DQI in benzene, the quintessential prototype for such effects in electron transport through molecules. We present our method for characterizing and analyzing DQI in Section \ref{sec:results}; this is the principal contribution of the present work. Section \ref{sec:discussion} then applies our framework to numerous examples, including benzene, anthracene derivatives, and cross-conjugated molecules. Finally, we summarize and conclude in Section \ref{sec:conclusions}.

\section{Background: Landauer-B\"uttiker Theory and Benzene}
\label{sec:example}
In this section we review DQI in a benzene molecule as described by a tight-binding model. This system has become the standard example of DQI in molecular electron transport, and a detailed analysis of it is presented in \cite{hansen-194704-2009}. Herein we summarize the pertinent details, which will provide context and an early example for the analysis we develop in Section \ref{sec:results}. A full description of this model can be found in Section \ref{sec:discussion}, the Supplemental Information, and \cite{hansen-194704-2009}. However, before we discuss electron transport through benzene, we must first introduce the Landauer-B\"uttiker theory for electron transport.

\subsection{Landauer-B\"uttiker Theory}
Within the limit of coherent scattering, electron transport through molecules is described by the Landauer-B\"uttiker formalism \cite{buttiker-6207-1985, imry-s306-1999, bk:cuevas-2010}. The transmission function, $T(E)$, is the key quantity, which essentially describes the probability that an electron with energy $E$ successfully tunnels from one electrode to the other through the molecule. In the limit of zero applied bias, the steady-state conductance through the electrode-molecule-electrode junction is
\[
G=\mathrm{G}_0 T(E_\mathrm{F}),
\]
where $\mathrm{G}_0\equiv 2e^2/h$ is the quantum of conductance and $E_\mathrm{F}$ is the Fermi energy of the junction. From a theoretical perspective, the transmission function is obtained from the Hamiltonian of the isolated molecule, $\mathbf{H}_0$, and self-energies, $\mathbf{\Sigma}_\mathrm{L/R}(E)$, that describe how the molecule couples to the left/right electrode. The self-energies are effectively open-system boundary conditions on the molecule. Then,
\begin{equation}
T(E) = \mathrm{Tr}\left[ \mathbf{G}(E) \mathbf{\Gamma}_\mathrm{L}(E) \mathbf{G}^\dagger(E) \mathbf{\Gamma}_\mathrm{R}(E) \right],
\label{eq:transmission}
\end{equation}
where
\[
\mathbf{G}(E) = \left[ E\mathbf{I} - \mathbf{H}_0 - \mathbf{\Sigma}_\mathrm{L}(E) - \mathbf{\Sigma}_\mathrm{R}(E)\right]^{-1}
\]
is the Green function of the molecule (as modified by the electrodes) and
\[
\mathbf{\Gamma}_\mathrm{L/R}(E) = i \left[ \mathbf{\Sigma}_\mathrm{L/R}(E) - \mathbf{\Sigma}_\mathrm{L/R}^\dagger(E) \right]
\]
is the spectral density for coupling the molecule to the left/right electrode. As a rough rule-of-thumb, the transmission function peaks at an energy $E$ if $E$ is the real part of an eigenvalue of $\mathbf{H}_0+\mathbf{\Sigma}_\mathrm{L}(E)+\mathbf{\Sigma}_\mathrm{R}(E)$; that is, there is a molecular resonance at $E$. Real eigenvalues indicate the presence of bound states (\textit{i.e.},\ molecular orbitals that do not couple to either electrode) in the molecular junction \cite{taylor-245407-2001, dhar-085119-2006}, which are inconsequential to steady-state transport and can be neglected. In what follows, we focus on the transmission function instead of conductance so that we can look at many possible behaviors with a only a few examples. The conductance can always be obtained by evaluating $T(E)$ at the Fermi energy.

It is commonly assumed that each electrode only couples to one site of the molecule---that there is only one conduction channel through the junction---such that DQI perfectly reflects electrons that enter the junction with energy $E$. $E$ is called the location of DQI. More mathematically, DQI at $E$ means the junction yields $T(E)=0$ when $\text{rank}(\mathbf{\Sigma}_\mathrm{L}(E))=1$ or $\text{rank}(\mathbf{\Sigma}_\mathrm{R}(E))=1$, where $\text{rank}(\mathbf{\Sigma}_\mathrm{L/R}(E))$ can be regarded as the number of ``bonds'' between the molecule and the left/right electrode. Identifying DQI is thus tantamount to finding roots of the transmission function \cite{bowen-2754-1995}. \cite{reuter-181103-2014} discusses an approach for describing DQI when there is more than one channel through the junction.

\subsection{Electron Transport through Benzene}
\begin{figure}
\resizebox{8.5cm}{!}{\includegraphics{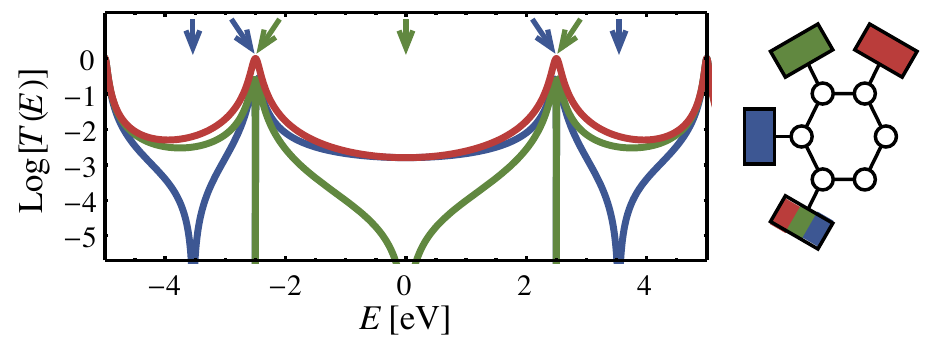}}
\caption{\label{fig:benzene-transmission}Transmission functions for electron transport through a benzene molecule with electrodes connected in the \textit{para} (red), \textit{meta} (green), and \textit{ortho} (blue) configurations. \textit{Meta}-benzene has the most robust (widest signature in $T(E)$) destructive interference effect, at $E=0$~eV, followed by the effects at $E\approx\pm3.5$~eV in \textit{ortho}-benzene, and by those at $E\approx\pm2.5$~eV in both \textit{ortho}- and \textit{meta}-benzene. This figure is modified, with permission, from \cite{reuter-181103-2014}, copyright 2014, AIP Publishing LLC.}
\end{figure}

Figure \ref{fig:benzene-transmission} shows the transmission functions for benzene connected to the electrodes in \textit{ortho}, \textit{meta}, and \textit{para} configurations. The arrows at the top of the figure show energies where at least one of the configurations exhibits DQI. It is clear that the existence and location of DQI depends on where the electrodes couple to the molecule. \textit{Ortho}-benzene shows 4 instances of DQI, \textit{meta}-benzene shows 3, and \textit{para}-benzene none.

A closer inspection of the transmission functions also reveals that not all instances of DQI are alike. The DQI at $E=0$~eV in \textit{meta}-benzene is very wide; that is, transmission is suppressed (although not 0) in a wide energy region around $E=0$~eV. Contrast this behavior with the DQI near $E=\pm2.5$~eV in both \textit{ortho}- and \textit{meta}-benzene. These effects are very narrow such that small changes in $E$ create large changes in transmission. Last is the DQI in \textit{ortho}-benzene near $E=\pm3.5$~eV, which is intermediate in width.

This width of DQI in the transmission function leads to the idea of ``robustness''. Suppose the junction's Fermi energy is close to $0$~eV in the \textit{meta}-benzene system. DQI is nearby; the transmission and thus conductance will be very low such that the DQI would be observable. This makes the DQI robust. On the other hand, if the Fermi energy were near $\pm2.5$~eV, the DQI might not be experimentally observed due to its narrow energy range. Such DQI is less robust.

Hansen \textit{et al}.\ \cite{hansen-194704-2009} further classified these instances of DQI in benzene as either ``multi-path'' or ``resonance''. Multi-path DQI stems from competing paths around the benzene molecule. The two paths essentially cancel each other through destructive interference, resulting in zero transmission. Contrasting, resonance DQI comes from the molecule's electronic structure alone, indicating substructure within $\mathbf{H}_0$. All DQI in the benzene configurations is multi-path except the instance through \textit{meta}-benzene at $E=0$~eV, which is resonance.

Some of the logic for distinguishing multi-path and resonance DQI in \cite{hansen-194704-2009} is inextricably linked to the cyclic structure of benzene. Knowing that acyclic molecules can also exhibit DQI \cite{solomon-17301-2008}, one of our goals in the present discussion is to generalize this classification. We will ultimately show in Sections \ref{sec:results} and \ref{sec:discussion} that resonance DQI (as generalized) is more robust than multi-path DQI. Each also has a distinct signature in the molecule's electronic structure. In this way, our characterization helps predict the experimental observability of DQI.

\section{Results}
\label{sec:results}
The benzene example in the previous section demonstrates the various types of DQI and, in the case of cyclic molecules, provides a scheme for classifying them. In this section we discuss DQI more broadly and generalize the characterization scheme. The following discussion is the primary contribution of this work.

Without making assumptions about the molecule or the molecule-electrode couplings (that is, without restricting our attention to conjugated hydrocarbons, tight-binding models, or junctions with a single conduction channel), DQI is described by a generalized eigenvalue problem \cite{reuter-181103-2014}. To reach this result, we must distinguish the parts of the molecule that directly couple to a particular electrode from the parts that do not. The kernels of $\mathbf{\Gamma}_\mathrm{L}(E)$ and $\mathbf{\Gamma}_\mathrm{R}(E)$, denoted $\text{Ker}[\mathbf{\Gamma}_\mathrm{L/R}(E)]$, accomplish this. In physical terms, $\text{Ker}[\mathbf{\Gamma}_\mathrm{L/R}(E)]$ is the set of all molecular state vectors (\textit{i.e.},\ kets) that do not directly couple to the left/right electrode. 

DQI appears at energies where a state vector that is decoupled from the left (right) electrode is unchanged by $\mathbf{H}_0+\mathbf{\Sigma}_\mathrm{L}(E)+\mathbf{\Sigma}_\mathrm{R}(E)$ in the molecular part that is decoupled from the right (left) electrode. The examples in Section \ref{sec:discussion} will help illustrate this idea. Mathematically, this condition is described by the  generalized eigenvalue problem
\[
\left[ \mathbf{H}_0 + \mathbf{\Sigma}_\mathrm{L}(E)+ \mathbf{\Sigma}_\mathrm{R}(E) \right]_{\text{Ker}[\mathbf{\Gamma}_\mathrm{R}(E)]\to\text{Ker}[\mathbf{\Gamma}_\mathrm{L}(E)]} \ket{\varphi_\mathrm{R}} = E \left(\mathbf{I}\right)_{\text{Ker}[\mathbf{\Gamma}_\mathrm{R}(E)]\to\text{Ker}[\mathbf{\Gamma}_\mathrm{L}(E)]} \ket{\varphi_\mathrm{R}},
\]
where $\mathbf{I}$ is the identity. The notation $\left(\mathbf{O}\right)_{A\to B}$ denotes an operator restriction where the operator $\mathbf{O}$ is only applied to state vectors from $A$ and the resulting state vectors are projected into $B$. In all but pathological cases, this equation can be simplified \footnote{Note that $\mathbf{\Gamma}_\mathrm{L/R}(E)$ is essentially the anti-Hermitian part of $\mathbf{\Sigma}_\mathrm{L/R}(E)$, such that only the Hermitian parts of $\mathbf{\Sigma}_\mathrm{L}(E)$ and $\mathbf{\Sigma}_\mathrm{R}(E)$ survive the restriction. Because the Hermitian and anti-Hermitian parts of $\mathbf{\Sigma}_\mathrm{L/R}(E)$ are related by a Kramers-Kronig relation (\textit{i.e.},\ Hilbert transform) \cite{bk:economou-2006}, the kernel of the Hermitian part will contain the kernel of the anti-Hermitian part except at rare and isolated energies. We also assume that $\text{Ker}[\mathbf{\Gamma}_\mathrm{L/R}(E)]$ does not depend on $E$, meaning that the parts of the molecule coupled to the electrodes do not change with $E$.} to%
\begin{subequations}
\label{eq:interference-eigprob}
\begin{equation}
\left( \mathbf{H}_0 \right)_{\text{Ker}[\mathbf{\Gamma}_\mathrm{R}]\to\text{Ker}[\mathbf{\Gamma}_\mathrm{L}]} \ket{\varphi_\mathrm{R}} = E \left(\mathbf{I}\right)_{\text{Ker}[\mathbf{\Gamma}_\mathrm{R}]\to\text{Ker}[\mathbf{\Gamma}_\mathrm{L}]} \ket{\varphi_\mathrm{R}}.
\end{equation}
We immediately see that DQI is primarily caused by substructure within the molecular Hamiltonian; the electrodes only serve to identify the parts of the molecule that are not coupled to each electrode. Finally, owing to the asymmetry between left and right electrodes, the left eigenvectors will be generally unrelated to the right eigenvectors, and are described by
\begin{equation}
\bra{\varphi_\mathrm{L}} \left( \mathbf{H}_0 \right)_{\text{Ker}[\mathbf{\Gamma}_\mathrm{R}]\to\text{Ker}[\mathbf{\Gamma}_\mathrm{L}]} = E \bra{\varphi_\mathrm{L}} \left( \mathbf{I} \right)_{\text{Ker}[\mathbf{\Gamma}_\mathrm{R}]\to\text{Ker}[\mathbf{\Gamma}_\mathrm{L}]}.
\end{equation}
\end{subequations}

Our previous work \cite{reuter-181103-2014} focused only on the existence and locations of DQI, where it was sufficient to find eigenvalues $E$ that satisfy Eq.\ \eqref{eq:interference-eigprob}. But Eq.\ \eqref{eq:interference-eigprob} also associates a left eigenvector $\ket{\varphi_\mathrm{L}}$ and a right eigenvector $\ket{\varphi_\mathrm{R}}$ with each instance of DQI. These eigenvectors, which we call ``interference vectors'' in the following discussion, provide the means to generalize the characterization scheme of Hansen \textit{et al}.\ \cite{hansen-194704-2009} from cyclic molecules to arbitrary molecules. Furthermore, each interference vector can be expanded in the molecular orbital basis, thereby revealing the participation of each molecular orbital in the DQI. Before proceeding, we note that the left and right interference vectors for the same instance of DQI are essentially unrelated to each other. This is a general property of generalized eigenvalue problems \cite{wilkinson-285-1979, van-dooren-103-1979}.

\begin{figure}
\resizebox{8.5cm}{!}{\includegraphics{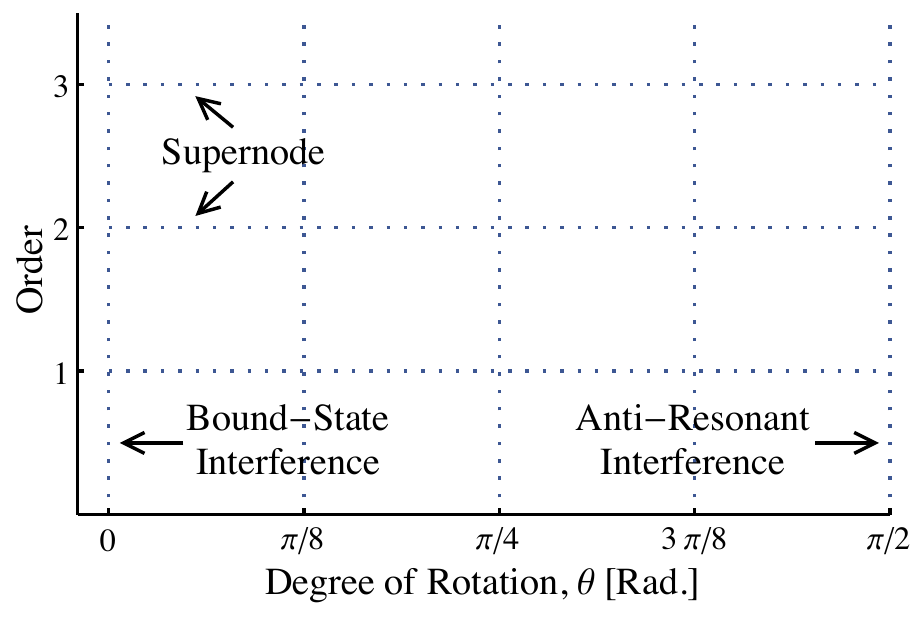}}
\caption{\label{fig:scheme}Graphical description of our characterization scheme for DQI. On the horizontal axis is the ``degree of rotation'' of DQI, $\theta$, given by Eq.\ \eqref{eq:rotation}. When $\theta=0$, the left and/or right interference vector is a molecular orbital and the DQI is classified as bound-state. When $\theta=\pi/2$, DQI results from an anti-resonance within the molecular Hamiltonian, generalizing the resonance-type DQI from \cite{hansen-194704-2009}. Intermediate degrees of rotation are also possible, evincing a continuous domain with bound-state and anti-resonant DQI as opposite extremes. On the vertical axis is the ``order'' of DQI, which describes the shape of the transmission function around the DQI. Most DQI is first-order, although higher orders have been theoretically predicted \cite{bergfield-5314-2010} and are called supernodes.}
\end{figure}

There are two key properties of the interference vectors and eigenvalues that lead to our generalized characterization scheme, which is graphically summarized in Figure \ref{fig:scheme}. We here define these properties and state the characterization scheme. Rationale and additional details will be presented alongside examples in Section \ref{sec:discussion}.

(i) The ``degree of rotation'' of the interference vectors, defined by
\begin{equation}
\theta = \sqrt{ \mathrm{arccos}\left(\frac{\left|\bok{\varphi_\mathrm{L}}{E\mathbf{I} - \mathbf{H}_0}{\varphi_\mathrm{L}}\right|}{\left\| \ket{\varphi_\mathrm{L}} \right\| \left\| \left( E\mathbf{I} -\mathbf{H}_0 \right) \ket{\varphi_\mathrm{L}} \right\|} \right) \mathrm{arccos}\left(\frac{\left|\bok{\varphi_\mathrm{R}}{E\mathbf{I} -\mathbf{H}_0}{\varphi_\mathrm{R}}\right|}{\left\| \ket{\varphi_\mathrm{R}} \right\| \left\| \left( E\mathbf{I} -\mathbf{H}_0 \right) \ket{\varphi_\mathrm{R}} \right\|} \right)}.
\label{eq:rotation}
\end{equation}
From vector calculus, either $\mathrm{arccos}$ expression in Eq.\ \eqref{eq:rotation} is the angle that the respective interference vector is rotated by $E\mathbf{I} - \mathbf{H}_0$ \cite{bk:gustafson-2012}, with this angle defined as 0 if the interference vector is an eigenvector of $\mathbf{H}_0$ for eigenvalue $E$. As we will see in the examples, $\theta$ relates to the robustness of DQI: $\theta=0$ produces a very narrow effect in the transmission function, whereas a larger $\theta$ (up to a maximum of $\theta=\pi/2$) creates wider effects in $T(E)$.

(ii) The ``order'' of the interference vector, which comes from the defectiveness of the DQI's eigenvalue in Eq.\ \eqref{eq:interference-eigprob}. (Recall from linear algebra that a degenerate eigenvalue may lack a complete set of linearly independent eigenvectors, in which case it is called defective. The Kronecker canonical form \cite{wilkinson-285-1979, van-dooren-103-1979} helps identify these cases.) The order predicts the shape of $T(E)$ around DQI located at $E_i$. If the order is $n=1,2,3,\ldots$, then a Taylor series expansion of $T(E)$ around $E_i$ has a leading term of $\mathcal{O}[(E-E_i)^{2n}]$. That is, the first $2n-1$ derivatives of $T(E)$ at $E_i$ are $0$. Second- and higher-order DQI have been previously discussed \cite{bergfield-5314-2010}; however, they are very sensitive to perturbation such that most DQI is first-order.

These two quantities, the degree of rotation $\theta\in[0,\pi/2]$ and the order $n\in\{1,2,3, \ldots\}$, constitute our characterization scheme for DQI. Both are readily obtained from Eq.\ \eqref{eq:interference-eigprob}: The locations and orders of all instances of DQI come from the eigenvalues of Eq.\ \eqref{eq:interference-eigprob} and their degeneracies, and the degrees of rotation from the corresponding left and right interference vectors. For some additional terminology,
\begin{itemize}
\item DQI that is not first-order corresponds to so-called supernodes \cite{bergfield-5314-2010}.
\item If one or both of the interference vectors are molecular orbitals such that $\theta=0$, the DQI is called ``bound-state''.
\item DQI is ``anti-resonant'' when $\theta=\pi/2$. In this case, both interference vectors are rotated $90^\circ$ by $E\mathbf{I}-\mathbf{H}_0$, meaning they can be regarded as molecular anti-resonances. Resonance-type DQI from \cite{hansen-194704-2009} fall into this category.
\item Intermediate values of $\theta$ are not given special names, but would belong to the multi-path class from \cite{hansen-194704-2009}.
\end{itemize}

\section{Discussion}
\label{sec:discussion}
We now present several example systems that showcase our characterization scheme for DQI. First is a simple model that exhibits the development and utility of the degree of rotation. We then proceed to more chemically-relevant examples, including benzene, cross-conjugated molecules, anthracene derivatives, and molecules with non-trivial couplings to the electrodes. Development of the order metric will be presented alongside the cross-conjugated molecules.

For simplicity, all of our examples employ tight-binding models, even though the theory and characterization scheme are more general. Unless noted otherwise, each ``atom'' in the molecule has a single orbital with an on-site energy of $\varepsilon=0$~eV and couples to its nearest neighbors with $\beta=-2.5$~eV. Finally, because the magnitude of electrode-molecule coupling is not germane to DQI [see Eq.\ \eqref{eq:interference-eigprob}], we invoke the wide-band limit \cite{verzijl-094102-2013} for the electrodes. Matrix elements for sites where the molecule couples to an electrode are $-i\Gamma=-0.1i$~eV. Full details about our models and computations can be found in the Supplemental Information.

\subsection{Three-Site Model}
\label{sec:discussion:3site}
\begin{figure}
\resizebox{8.5cm}{!}{\includegraphics{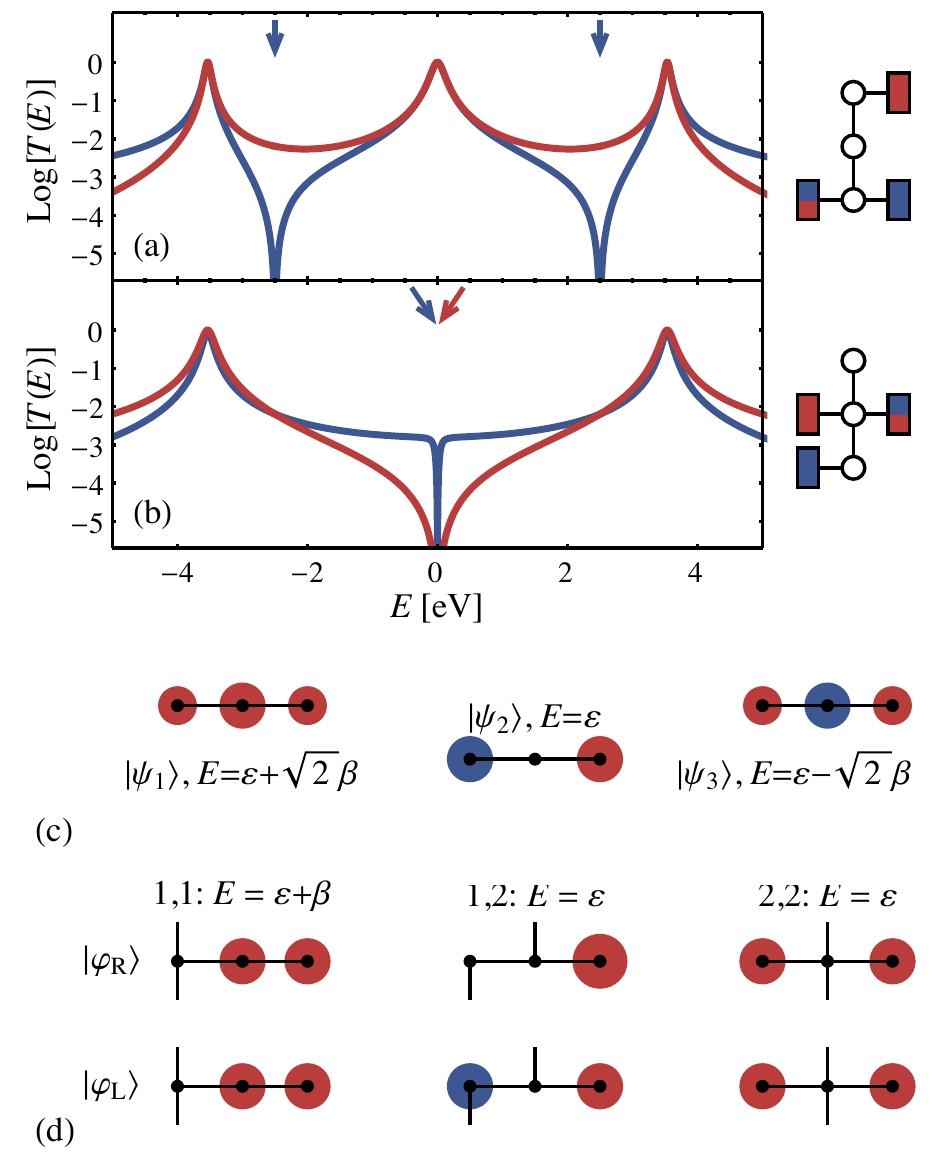}}
\caption{\label{fig:3site}DQI in a three-site tight-binding model. (a) Transmission functions when the electrodes are connected at the 1,1 (blue) and 1,3 (red) positions. There are two instances of DQI at $\varepsilon\pm\beta=\pm2.5$~eV in the 1,1 case. (b) Transmission functions when the electrodes are connected at the 1,2 (blue) and 2,2 (red) positions. Both of these configurations exhibit DQI at $\varepsilon=0$~eV. (c) Molecular orbitals of the bare molecule along with their energies. Red and blue ``clouds'' represent the phase and magnitude of the wavefunction. (d) Left and right interference vectors for three instances of DQI. The DQI at $E=\varepsilon-\beta=2.5$~eV in the 1,1 configuration is omitted because it is very similar to that at $E=\varepsilon+\beta=-2.5$~eV. All instances of DQI have a degree of rotation of $\theta=\pi/2$ (they are anti-resonant DQI) except the one in the 1,2 configuration, which is bound-state DQI ($\theta=0$). Details on the calculations can be found in the Supplemental Information.}
\end{figure}

Our opening example is a three-site tight-binding model, as pictured in Figure \ref{fig:3site}. Although this model might be a representation of propene, we do not place any physical or chemical significance on the results. Rather, this simple example is intended to motivate the degree of rotation and to demonstrate the different types of DQI that can occur in more physically-meaningful systems.

Figures \ref{fig:3site}(a) and (b) display the transmission functions when the electrodes are attached to the molecule in the four symmetrically-distinct configurations. The 1,3 configuration does not result in any DQI, whereas the 1,2 and 2,2 configurations each have 1 instance of DQI at $E=\varepsilon=0$~eV and the 1,1 configuration has 2 instances of DQI at $E=\varepsilon\pm\beta=\pm2.5$~eV. Qualitatively, DQI in the 2,2 configuration is the most robust (it has the widest valley in the transmission function), and DQI in the 1,2 configuration is the least robust.

We now develop the degree of rotation defined in Eq.\ \eqref{eq:rotation}. The first step is to examine the molecular orbitals (MOs) of the three-site model [Figure \ref{fig:3site}(c)], with a focus on their nodal structure. The middle MO, labeled $\ket{\psi_2}$ in the figure, has a node on the middle site. We can immediately draw several conclusions about the transport properties of this molecule when connected in some configurations. Suppose both electrodes couple to this middle site (\textit{i.e.},\ the 2,2 configuration). There will be a bound state \cite{taylor-245407-2001, dhar-085119-2006} in the junction because the MO $\ket{\psi_2}$ does not couple to either electrode and thus does not participate in transport. If we move exactly one electrode so that it couples to a different site (\textit{e.g.},\ the 1,2 configuration), then there will be DQI at that orbital's energy. The MO is still decoupled from one electrode, and Figure \ref{fig:3site}(b) verifies a narrow instance of DQI at $E=\varepsilon=0$~eV. Furthermore, as shown in Figure \ref{fig:3site}(d), one of the interference vectors for this DQI is the MO itself. Because this interference vector is a MO that forms a bound state \textit{in a different molecule-electrode configuration}, we call this ``bound-state'' DQI. Knowledge of the transport through one configuration can provide information on other configurations.

The nodal structure of the interference vectors is also important for characterizing DQI. By construction in Eq.\ \eqref{eq:interference-eigprob}, each interference vector will be decoupled from at least one electrode (having a node at those sites), with corresponding left and right interference vectors guaranteed to be decoupled from opposite electrodes. For the DQI at $E=\varepsilon$ in the 1,2 configuration, the left interference vector is decoupled from the electrode at site 2, but not from the electrode at site 1 [see Figure \ref{fig:3site}(d)]. The corresponding right interference vector is decoupled from both electrodes. This structure in how the interference vectors decouples from the electrodes relates to robustness of DQI. Suppose that one interference vector is only decoupled from one electrode. Perturbations to the system are likely to recouple the interference vector to the electrode, thereby eliminating the DQI. In contrast, a vector that is decoupled from both electrodes will require more perturbation to recouple the molecule, resulting in more robust DQI.

To help demonstrate this idea, consider the DQI at $E=\varepsilon$ in the 2,2 configuration. Both the left and right interference vectors are the same, and neither is coupled to either electrode [Figure \ref{fig:3site}(d)]. As we might expect from the above discussion, this instance of DQI is much more robust than that in the 1,2 configuration. We also see that the interference vectors are not MOs; instead, $\ket{\varphi_\mathrm{L}}=\ket{\varphi_\mathrm{R}}=(\ket{\psi_1}+\ket{\psi_3})/\sqrt{2}$. From our definition, this cannot be bound-state DQI.

The degree of rotation $\theta$ quantifies this nodal structure within the interference vectors by examining geometric properties of the interference vectors. Specifically, we look at the rotation of an interference vector by $E\mathbf{I}-\mathbf{H}_0$, where $E$ is the location of the specific instance of DQI and the angle of rotation is \cite{bk:gustafson-2012}
\[
\mathrm{arccos}\left( \frac{\left| \bok{\varphi_\mathrm{L/R}}{E\mathbf{I}-\mathbf{H}_0}{\varphi_\mathrm{L/R}} \right|}{\left\| \ket{\varphi_\mathrm{L/R}} \right\| \left\| \left( E\mathbf{I} - \mathbf{H}_0 \right) \ket{\varphi_\mathrm{L/R}} \right\|} \right),
\]
assuming $\ket{\varphi_\mathrm{L/R}}$ is neither the zero vector nor an eigenvector of $E\mathbf{I} - \mathbf{H}_0$ (in which case the angle is defined to be 0). In the case of bound-state DQI, at least one of the interference vectors is a MO---that is, it is an eigenvector of both $\mathbf{H}_0$ and $E\mathbf{I}-\mathbf{H}_0$---such that it is not rotated by $E\mathbf{I}-\mathbf{H}_0$. Thus, $\ket{\varphi_\mathrm{L}}$ or $\ket{\varphi_\mathrm{R}}$ having a rotation of $0$ indicates bound-state DQI. In contrast, the interference vectors for DQI at $E=\varepsilon$ in the 2,2 configuration are rotated $\pi/2$ by $E\mathbf{I}-\mathbf{H}_0$; that is, $(E\mathbf{I}-\mathbf{H}_0)\ket{\varphi_\mathrm{L/R}}$ is orthogonal to $\ket{\varphi_\mathrm{L/R}}$.

It is straightforward to show that an interference vector that is decoupled from both electrodes will be rotated $\pi/2$ by $E\mathbf{I}-\mathbf{H}_0$. Because of the nodes at both electrodes, it will also be rotated $\pi/2$ by $E\mathbf{I}-\mathbf{H}_0-\mathbf{\Sigma}_\mathrm{L}(E)-\mathbf{\Sigma}_\mathrm{R}(E)$. Given that eigenvectors of $E\mathbf{I}-\mathbf{H}_0-\mathbf{\Sigma}_\mathrm{L}(E)-\mathbf{\Sigma}_\mathrm{R}(E)$ correspond to molecular resonances in the junction, there's a notational appeal to classifying these interference vectors as anti-resonant DQI. The caveat, as can be seen in the bound-state DQI from our three-site example, is that \textit{both} the left and right interference vectors must be rotated by $\pi/2$ for the instance of DQI to be considered anti-resonant.

We finally arrive at the degree of rotation $\theta$ defined in Eq.\ \eqref{eq:rotation}. DQI is bound-state if either of its interference vectors is rotated 0 by $E\mathbf{I}-\mathbf{H}_0$ and, similarly, is anti-resonant if both of its interference vectors are rotated $\pi/2$. The product of the angles for the left and right interference vectors in Eq.\ \eqref{eq:rotation} accomplishes this. Thus, $0\le\theta\le\pi/2$, where $\theta=0$ is bound-state DQI and $\theta=\pi/2$ is anti-resonant DQI. Higher values of $\theta$ correspond to more robust instances of DQI. Note that the two instances of DQI in the 1,1 configuration of the three-site model are also anti-resonant DQI, such that they do not advance the discussion. Other examples (see below) will show DQI with intermediate values of $\theta$.

Finally, before moving on to the next example, we discuss one other application for the degree of rotation. In addition to quantifying the nodal structure of the interference vectors of DQI, it also reveals some insight into the steepness of the transmission function around the DQI. Consider the Taylor series of $T(E)$ around the location of each instance of DQI in the three-site model. For the bound-state DQI in the 1,2 configuration,
\[
T_{1,2}\left(E\approx \varepsilon\right) = \frac{4}{\beta^2}(E-\varepsilon)^2 + \mathcal{O}[(E-\varepsilon)^4].
\]
The leading term of the transmission function is quadratic (indicating first-order DQI), and its coefficient does not depend on $\Gamma$, the molecule-electrode coupling strength. In a likewise fashion, the transmission function around the anti-resonant DQI in the 2,2 configuration is
\[
T_{2,2}\left(E\approx \varepsilon\right) = \frac{\Gamma^2}{\beta^4}(E-\varepsilon)^2 + \mathcal{O}[(E-\varepsilon)^4]
\]
and for either anti-resonant DQI in the 1,1 configuration,
\[
T_{1,1}\left(E\approx \varepsilon\pm\beta\right) = \frac{16\Gamma^2}{\beta^4}\left( E - \left(\varepsilon\pm\beta\right)\right)^2 + \mathcal{O}\left[ \left( E- \left( \varepsilon\pm\beta \right) \right)^3 \right].
\]
These leading coefficients now contain both $\Gamma$ and $\beta$. When combined with other examples (see below), it appears that $\Gamma$ and $\beta$ are competing factors in these leading coefficients. One of them is missing in bound-state DQI, and they are ``in-phase'' with each other in anti-resonant DQI. That is, they essentially appear as ratios. As we will see in the next example, intermediate DQI will have them be ``out-of-phase''.

\subsection{Benzene}
\label{sec:discussion:benzene}
No characterization of DQI in molecules would be complete without showcasing benzene. We briefly discussed this system in Section \ref{sec:example}, and Figure \ref{fig:benzene-transmission} shows the transmission functions for benzene connected in the \textit{ortho}, \textit{meta}, and \textit{para} configurations. In the end, the analysis of DQI in benzene is very similar to that of the three-site model in the previous example. The only complicating factor is the degeneracy of benzene's highest-occupied MO and lowest-unoccupied MO. As we will now see, this issue is straightforwardly handled.

\begin{figure}
\resizebox{8.5cm}{!}{\includegraphics{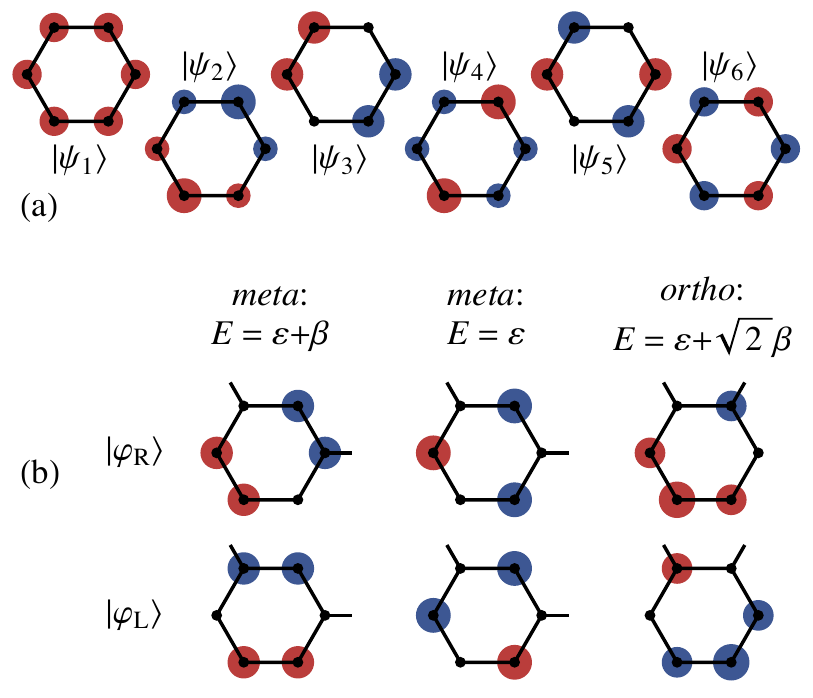}}
\caption{\label{fig:benzene-ivs}(a) Molecular orbitals of a bare benzene molecule. $\ket{\psi_1}$ has an energy of $\varepsilon+2\beta$, $\ket{\psi_2}$ and $\ket{\psi_3}$ are degenerate with an energy of $\varepsilon+\beta$, $\ket{\psi_4}$ and $\ket{\psi_5}$ are degenerate with an energy of $\varepsilon-\beta$, and $\ket{\psi_6}$ has energy $\varepsilon-2\beta$. (b) Interference vectors for three instances of DQI in benzene. The instances of DQI at $E=\varepsilon\pm\beta$ in the \textit{meta} and \textit{ortho} configurations are bound-state ($\theta=0$), the instance at $E=\varepsilon$ in the \textit{meta} configuration is anti-resonant ($\theta=\pi/2$), and the instances at $E=\varepsilon\pm\sqrt{2}\beta$ in the \textit{ortho} configuration are intermediate ($\theta=1.11$). All DQI in benzene is first-order.}
\end{figure}

Similar to our analysis of DQI in the three-site model, we begin with an examination of benzene's MOs, which are depicted in Figure \ref{fig:benzene-ivs}(a). The MOs labeled $\ket{\psi_3}$ and $\ket{\psi_5}$ each have two nodes such that the \textit{para} configuration will have bound states at $E=\varepsilon\pm\beta=\mp2.5$~eV. Moving exactly one electrode off of these nodes produces a system in either the \textit{ortho} or \textit{meta} configuration. Such a system should exhibit bound-state DQI at the energies of $\ket{\psi_3}$ ($\varepsilon+\beta=-2.5$~eV) and $\ket{\psi_5}$ ($\varepsilon-\beta=2.5$~eV), which is verified in Figure \ref{fig:benzene-transmission}. Figure \ref{fig:benzene-ivs}(b) then shows the interference vectors for one of these cases. As we would expect, the interference vectors are MOs (thus having $\theta=0$); however, they do not exactly match $\ket{\psi_3}$ or $\ket{\psi_5}$. Instead, the interference vectors are linear combinations either of $\ket{\psi_2}$ and $\ket{\psi_3}$ or of $\ket{\psi_4}$ and $\ket{\psi_5}$, which are still eigenvectors of the molecular Hamiltonian due to degeneracy of its respective eigenvalues.

In addition, we see that the DQI at $E=\varepsilon=0$~eV in \textit{meta} configuration is anti-resonant. Its interference vectors are displayed in Figure \ref{fig:benzene-ivs}(b), and are both decoupled from both electrodes. Finally, the DQI at $E=\varepsilon\pm\sqrt{2}\beta$ in the \textit{ortho} configuration provides our first example of DQI that is neither bound-state nor anti-resonant. These instances of DQI have $\theta=1.11$. As we might then expect, its interference vectors [Figure \ref{fig:benzene-ivs}(b)] are not MOs and are only decoupled from one electrode. An inspection of the transmission functions in Figure \ref{fig:benzene-transmission} reveals that they are also of intermediate robustness. All DQI in these benzene configurations are first-order.

As the final point of this example, we examine the Taylor series expansions of $T(E)$ around each instance of DQI. The anti-resonant DQI in \textit{meta}-benzene has
\[
T_\mathrm{meta}(E\approx\varepsilon) = \frac{\Gamma^2}{4\beta^4}\left( E-\varepsilon \right)^2 + \mathcal{O}\left[ \left( E - \varepsilon \right)^4 \right],
\]
which is still quadratic and, as we expect for anti-resonant DQI, has $\beta$ and $\Gamma$ in-phase with each other. The bound-state DQI in \textit{ortho}- or \textit{meta}-benzene produce
\[
T_\mathrm{ortho}(E\approx\varepsilon \pm \beta) = T_\mathrm{meta}(E\approx\varepsilon \pm \beta) = \frac{16}{\Gamma^2}\left( E-(\varepsilon\pm\beta) \right)^2 + \mathcal{O}\left[ \left( E - (\varepsilon\pm\beta) \right)^3 \right].
\]
$\beta$ is now missing from the coefficients, again supporting our classification as bound-state DQI. Last, the intermediate DQI in \text{ortho}-benzene has
\[
T_\mathrm{ortho}(E\approx\varepsilon \pm \sqrt{2} \beta) = \frac{32\Gamma^2}{(2\beta^2+\Gamma^2)^2}\left( E-(\varepsilon\pm\sqrt{2}\beta) \right)^2 + \mathcal{O}\left[ \left( E - (\varepsilon\pm\sqrt{2}\beta) \right)^3 \right].
\]
Both $\Gamma$ and $\beta$ are present in the coefficient, and its denominator suggests they are out-of-phase with each other.

\subsection{Cross-Conjugated Molecules: Order of DQI}
\label{sec:discussion:cc}
\begin{figure}
\resizebox{8.5cm}{!}{\includegraphics{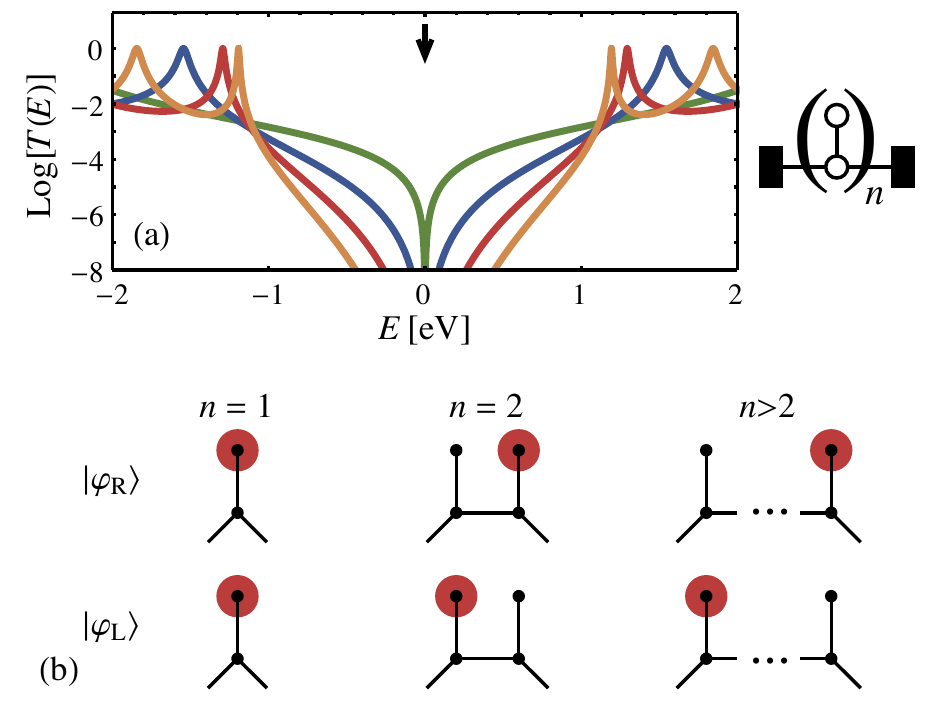}}
\caption{\label{fig:crossconj}(a) Transmission functions for ``comb'' oligomers with $n=1$ (green), $2$ (blue), $3$ (red), and $4$ (orange). The DQI at $E=0$~eV is $n$th-order, meaning the transmission function is $\mathcal{O}[(E-0)^{2n}]$ near the DQI. Higher-order DQI (resulting in supernodes) is more robust than lower-order DQI, but is also rarer than lower-order DQI (as discussed in the main text). (b) Interference vectors for DQI in these systems with $n=1$, $n=2$, and general $n$. All DQI in these comb oligomers is anti-resonant.}
\end{figure}

All of our examples so far have demonstrated first-order DQI. We now discuss higher-order DQI with the ``comb'' molecules of \cite{fowler-174708-2009}, which may be coarse-grained representations of cross-conjugated oligomers \cite{andrews-17309-2008}. These molecules are schematically depicted in Figure \ref{fig:crossconj}(a). As mentioned in Section \ref{sec:results}, the order of DQI comes from the geometric degeneracy of the eigenvalue in Eq.\ \eqref{eq:interference-eigprob} and ultimately relates to the shape of the transmission function around the DQI. Specifically, $n$th-order DQI at $E_i$ produces
\[
T(E\approx E_i) = C \left( E - E_i \right)^{2n} + \mathcal{O}\left[ \left( E-E_i \right)^{2n+1} \right],
\]
where $C$ is a constant that depends on the molecule-electrode coupling strength ($\Gamma$) and/or molecular structure (\textit{e.g.},\ $\beta$ in our tight-binding models).

Figure \ref{fig:crossconj}(a) displays the transmission functions for the comb molecules with $n=1$ to $n=4$. All of these molecules have anti-resonant DQI at $E=\varepsilon=0$~eV [see the interference vectors in Figure \ref{fig:crossconj}(b)] with more robust DQI as $n$ increases. Because $\theta=\pi/2$ for all cases, we need another metric to describe this change in robustness. Appealing to Eq.\ \eqref{eq:interference-eigprob}, the comb molecule with $n$ repeat units produces a generalized eigenvalue problem that has $E=\varepsilon$ as an $n$-degenerate eigenvalue. However, there is only one left (right) interference vector for this eigenvalue regardless of $n$, indicating defectiveness in the eigenvalue problem. This level of defectiveness is the order of DQI.

When $n=1$, the eigenvalue is not defective; $E=\varepsilon$ is a simple eigenvalue of Eq.\ \eqref{eq:interference-eigprob} with a single (left or right) eigenvector. The DQI is thus first-order and
\[
T_{n=1}\left(E\approx\varepsilon\right) = \frac{4\Gamma^2}{\beta^4} \left(E-\varepsilon\right)^2 + \mathcal{O}\left[ \left( E - \varepsilon \right)^4 \right].
\]
Consistent with anti-resonant DQI, the leading coefficient in the Taylor expansion is an ``in-phase'' relationship between the coupling strength and the molecule structure. If we increase $n$ to $2$, $E=\varepsilon$ becomes a doubly-degenerate eigenvalue, but it still has a single (left or right) interference vector. In linear algebra terms, the eigenvalue is defective. The DQI remains anti-resonant, but the shape of the transmission function around the DQI changes to
\[
T_{n=2}\left(E\approx\varepsilon\right) = \frac{4\Gamma^2}{\beta^6} \left(E-\varepsilon\right)^4 + \mathcal{O}\left[ \left( E - \varepsilon \right)^6 \right].
\]
This trend continues as the number of monomers (``teeth'') in the comb molecule increases. The DQI maintains a single (left or right) interference vector, but its degeneracy and order increase. In turn, the transmission function becomes flatter, and thus more robust, around the DQI:
\[
T_{n>2}\left(E\approx\varepsilon\right) = \frac{4\Gamma^2}{\beta^{2n+2}} \left(E-\varepsilon\right)^{2n} + \mathcal{O}\left[ \left( E - \varepsilon \right)^{2n+2} \right].
\]
This result is evident in Figure \ref{fig:crossconj}(a).

Polymerization generally increases the order of DQI if the monomer exhibits DQI and the monomers are suitably connected \cite{andrews-17309-2008, fowler-174708-2009, bergfield-5314-2010}. In this sense, polymerization could be one means to designing systems with more robust DQI; others are discussed in \cite{bergfield-5314-2010}. We note, however, that high-order DQI is very sensitive to perturbation and is thus rare \cite{bergfield-5314-2010}. Any form of disorder, perhaps due to molecular vibrations, will likely reduce the DQI to first-order. The DQI can still be relatively more robust; the transmission function will become quadratic around the DQI but with a small coefficient compared to higher-order terms.

As a mathematical aside, the order of DQI is the size of the Kronecker block for the DQI in the Kronecker canonical form \cite{wilkinson-285-1979, van-dooren-103-1979} of Eq.\ \eqref{eq:interference-eigprob}. Recall that standard eigenvalue problems are characterized by a diagonal decomposition or, more generally, the Jordan normal form if the matrix is not diagonalizable. The Kronecker canonical form extends this idea to generalized eigenvalue problems. Each instance of DQI will have a block in the Kronecker canonical form (similar to a Jordan block in the Jordan normal form), and the size of this block is the DQI order. The Kronecker canonical form also displays the general independence (\textit{i.e.},\ lack of relationship) between the left and right interference vectors.

\subsection{Anthracene Derivatives: Complex Energies}
\label{sec:discussion:aq}
Having developed and described our characterization scheme for DQI, we now apply it to two physical setups that are not well understood. First is DQI at complex energies \cite{bowen-2754-1995, reuter-181103-2014, pedersen-26919-2015}, and then nontrivial molecule-electrode couplings \cite{tsuji-224311-2014, hansen-6295-2016} in the next subsection.

\begin{figure}
\resizebox{8.5cm}{!}{\includegraphics{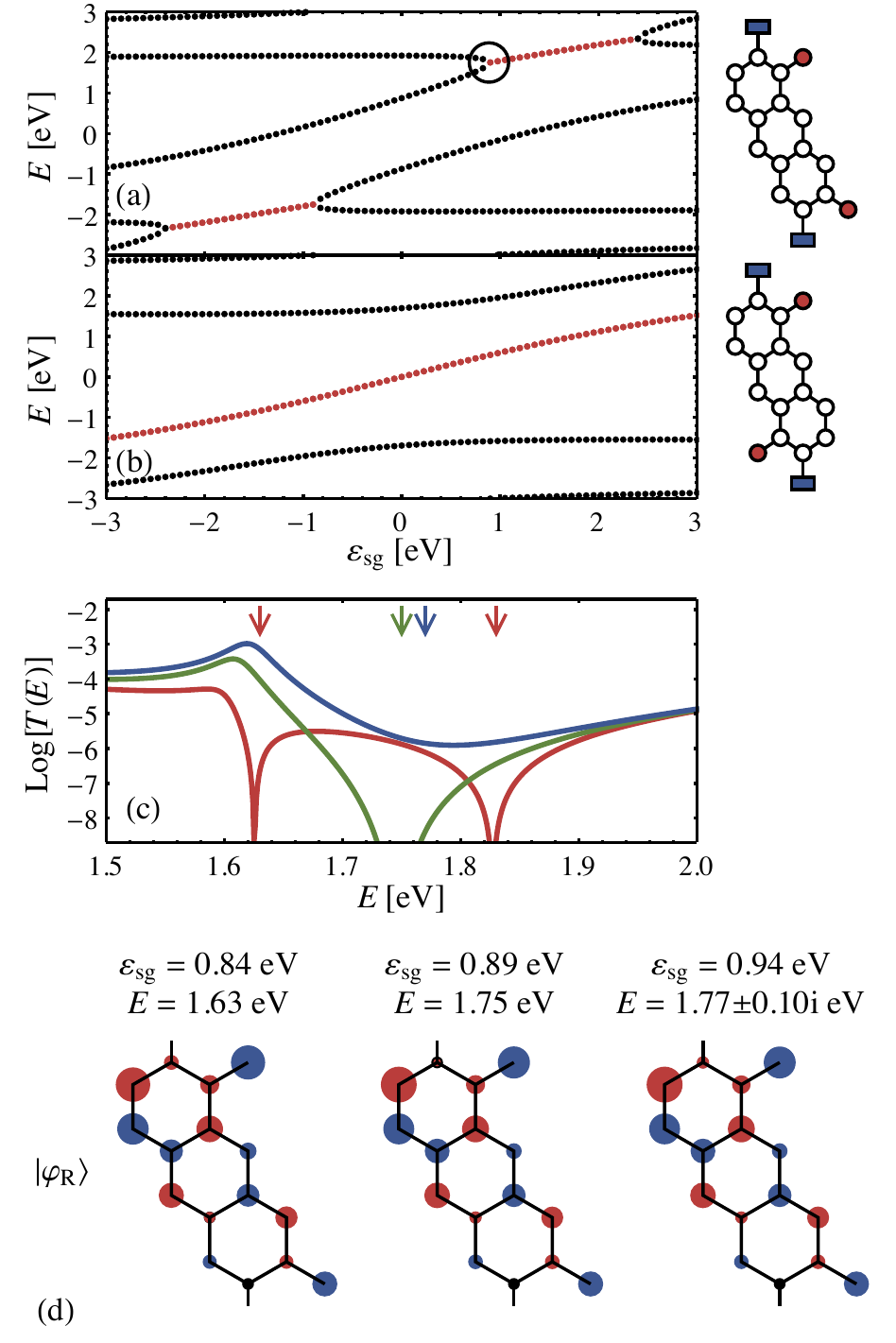}}
\caption{\label{fig:aq}DQI locations through anthracene derivatives. The blue rectangles in the schematics represent electrodes and the red circles are heteroatoms, which possess different site energies ($\varepsilon_\mathrm{sg}$) from the carbon atoms. (a) and (b) Locations of DQI for the depicted molecules as $\varepsilon_\mathrm{sg}$ is varied. Black dots denote real DQI energies, whereas red dots indicate the real parts of complex DQI energies. (c) Transmission functions for three values of $\varepsilon_\mathrm{sg}$ in the circled region of panel (a). Red: $\varepsilon_\mathrm{sg}=0.84$~eV. There are two distinct, first-order DQI instances at $E=1.63$~eV and $E=1.83$~eV. Green: $\varepsilon_\mathrm{sg}=0.89$~eV, where there is second-order DQI at $E=1.75$~eV. Blue: $\varepsilon_\mathrm{sg}=0.94$~eV; there are two instances of first-order DQI with complex conjugate energies, $\mathrm{Re}(E)=1.77$~eV. (d) Right interference vectors for the instances of DQI in panel (c). The first-order DQI at $E=1.83$~eV is similar to that at $E=1.63$~eV and is not pictured. Despite the differences in transmission (conductance), DQI at complex energies is chemically similar to DQI at real energies.}
\end{figure}

It was shown in \cite{bowen-2754-1995, reuter-181103-2014} that DQI can occur at complex energies. The transmission function does not drop to zero along the real axis in such an event, but instead exhibits a minimum near the real part of the complex energy. The blue curve in Figure \ref{fig:aq}(c) shows an example of transmission around complex DQI energies. In general, DQI at a complex energy occurs when two instances of DQI with real energies ``collide'' and the locations travel off into the complex plane \cite{bowen-2754-1995}. In this way, if a molecule exhibits DQI at two nearby energies, slight perturbations may cause the DQI to move off into the complex plane, thus increasing transmission \cite{pedersen-26919-2015}.

We use tight-binding models of anthracene derivatives from \cite{markussen-14311-2011} to demonstrate this effect. As schematically depicted in Figures \ref{fig:aq}(a) and (b), the two oxygen atoms from anthraquinone are placed at various positions around the anthracene structure and are considered to be generic heteroatoms. Within the tight-binding model, these two heteroatoms have on-site energies of $\varepsilon_\mathrm{sg}$ and couple to their neighboring carbon atoms with the same $\beta$ as the carbon-carbon bonds. Varying these on-site energies moves the locations of DQI, as displayed in Figures \ref{fig:aq}(a) and (b) for two configurations.

Most prominently, we see that changes in $\varepsilon_\mathrm{sg}$ do not change the number of instances of DQI (note that the complex energies come in conjugate pairs). In some cases [Figure \ref{fig:aq}(a)], DQI at complex energies emerge after two instances of DQI collide. In other cases [Figure \ref{fig:aq}(b)], there can be a ``band'' of DQI at complex energies.

To gain more insight into the nature of DQI at complex energies, we examine DQI near and at the circled ``collision'' in Figure \ref{fig:aq}(a). The transmission functions around this DQI for $\varepsilon_\mathrm{sg}=0.84$~eV, $\varepsilon_\mathrm{sg}=0.89$~eV, and $\varepsilon_\mathrm{sg}=0.94$~eV are displayed in Figure \ref{fig:aq}(c). Right interference vectors for DQI at these values of $\varepsilon_\mathrm{sg}$ are also displayed in Figure \ref{fig:aq}(d). In the first of these three cases ($\varepsilon_\mathrm{sg}=0.84$~eV), there are two distinct instances of first-order DQI at $E=1.63$~eV (degree of rotation $\theta=1.30$) and at $E=1.83$~eV ($\theta=1.34$). As $\varepsilon_\mathrm{sg}$ increases from this value, the locations of these instances of DQI approach each other and the interference vectors become more similar. Eventually, at $\varepsilon_\mathrm{sg}=0.89$~eV, the DQI locations collide and the interference vectors become identical. There is a single, second-order instance of DQI with $\theta=1.33$ at $E=1.75$~eV. As expected from its increased order and similar degree of rotation, this DQI is more robust than the instances of DQI in the previous case. Finally, the second-order DQI splits back into two first-order instances of DQI ($\theta=1.32$) as $\varepsilon_\mathrm{sg}$ increases further to $0.94$~eV. However, these instances of DQI now occur at complex conjugate energies. Because the DQI is not on the real axis, the transmission increases.

In the end, an inspection of the interference vectors in Figure \ref{fig:aq}(d) shows very few changes with $\varepsilon_\mathrm{sg}$. There are certainly small quantitative differences (see the Supplemental Information), but nothing of qualitative note. Although DQI at complex energies may seem chemically bizarre at first, there do not seem to be large chemical signatures for them when compared to DQI at real energies. Consequently, DQI at complex energies are probably of more applied value than fundamental. For example, if the on-site energies of the heteroatoms in our anthracene examples were gated, the DQI could produce effective transistors as the DQI is shifted on to or off of the experimentally-relevant real axis by the gate.

\subsection{Nontrivial Molecule-Electrode Couplings}
\label{sec:discussion:multicouple}
Our final example looks at systems with nontrivial molecule-electrode couplings \cite{reuter-181103-2014, tsuji-224311-2014, hansen-6295-2016}. Most studies of DQI use simple couplings, where each electrode couples to exactly one site of the tight-binding model. The various theories for DQI are built on this setup. In contrast, the generalized eigenvalue problem in Eq.\ \eqref{eq:interference-eigprob} places no limitations on the molecule-electrode coupling. Therefore, our present definition of interference vectors and characterization scheme for DQI is readily applicable.

\begin{figure}
\resizebox{8.5cm}{!}{\includegraphics{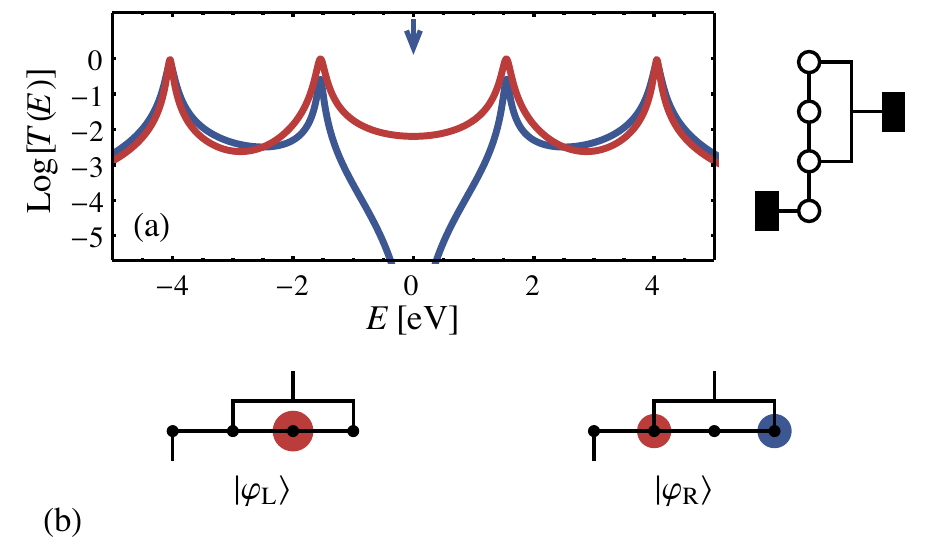}}
\caption{\label{fig:mc}DQI in a butadiene molecule that is nontrivially coupled to the electrodes. One electrode trivially couples to one site, whereas the other electrode couples to two other molecular sites. (a) Transmission functions for the case of incoherent (red) and coherent (blue) molecule-electrode coupling [Eqs.\ \eqref{eq:incoherent} and \eqref{eq:coherent}, respectively]. Only the coherently-coupled case exhibits DQI, with second-order DQI ($\theta=\pi/2$) at $E=\varepsilon=0$~eV. There is no DQI with incoherent coupling. (b) Left and right interference vectors for the DQI exhibited with coherent coupling.}
\end{figure}

The chief subtlety with nontrivial molecule-electrode couplings is that it becomes insufficient to only specify the molecular sites that couple to each electrode. Consider the tight-binding model for butadiene in Figure \ref{fig:mc}. As depicted, one electrode couples to a single site (enumerated as site 1) but the other couples to two sites (sites 2 and 4). Similar systems are discussed in \cite{tsuji-224311-2014}. The self-energy for coupling to the second electrode might appear as%
\begin{subequations}
\begin{equation}
\mathbf{\Sigma}(E) = -\frac{i\Gamma}{2} \left( \kb{\chi_2}{\chi_2} + \kb{\chi_4}{\chi_4} \right),
\label{eq:incoherent}
\end{equation}
where $\ket{\chi_j}$ is the AO on the $j$th site. On the other hand, the self-energy may also be
\begin{equation}
\mathbf{\Sigma}(E) = -\frac{i\Gamma}{2} \left( \kb{\chi_2}{\chi_2} + \kb{\chi_2}{\chi_4} + \kb{\chi_4}{\chi_4} + \kb{\chi_4}{\chi_2} \right).
\label{eq:coherent}
\end{equation}
\end{subequations}
In both cases, the electrode only couples to sites 2 and 4 of the molecule, but the first self-energy has a rank of 2 and the second has a rank of 1. (Recall that the rank of the self-energy can be loosely interpreted as the number of ``bonds'' between the electrode and the molecule.)

Hansen and Solomon \cite{hansen-6295-2016} refer to these couplings as incoherent and coherent, respectively. Such incoherent coupling essentially means that each molecular site independently interacts with the electrode, whereas the two sites' interactions are coordinated in the coherent case. This seemingly-small distinction has significant effects on the transmission function and on DQI in the molecule, as displayed in Figure \ref{fig:mc}(a). Second-order, anti-resonant DQI is present at $E=\varepsilon=0$~eV when the molecule is coherently coupled to the electrodes. There is no DQI, not even at complex energies, when the molecule is incoherently coupled to the electrodes.

This example highlights two points. First, nontrivial molecule-electrode couplings are more nuanced than the usual cases where each electrode couples to a single molecular site. Coherent coupling can lead to qualitatively different transport properties from incoherent coupling. Second, our characterization of DQI readily generalizes to cases of nontrivial coupling. The generalized eigenvalue problem in Eq.\ \eqref{eq:interference-eigprob} makes no assumptions about the style of coupling---only requiring $\mathrm{Ker}(\mathbf{\Gamma}_\mathrm{L/R})$---and provides interference vectors that can be analyzed in an identical fashion. Accordingly, Figure \ref{fig:mc}(b) shows the interference vectors for the coherently-coupled butadiene molecule.

\section{Conclusions}
\label{sec:conclusions}
In this work we developed a characterization scheme for DQI in electron transport through molecules. DQI is generally described by a generalized eigenvalue problem \cite{reuter-181103-2014}, Eq.\ \eqref{eq:interference-eigprob}, which also associates eigenvectors with DQI. These ``interference vectors'' are the basis for our DQI characterization scheme. On one hand, the interference vectors can be decomposed in the MO basis, thereby revealing the participation of each MO in DQI. On the other hand, they also have geometric properties that predict the robustness of DQI. We specifically analyzed two of these properties, order and degree of rotation, which form the basis of our characterization scheme. The order describes the shape of the transmission function around the DQI and the degree of rotation quantifies the nodal structure of the interference vectors where the electrodes couple to the molecule. Increased order and degree of rotation both lead to more robust DQI.

We then explored the utility of interference vectors and this characterization scheme with several model systems. For example, we found that DQI at a complex energy appears to be of more applied interest than fundamental. As the locations of DQI move into the complex plane, the interference vectors remained essentially unchanged. In the end, this style of analysis places DQI on a similar footing with more common analyses of molecular resonances in transport. Both the peaks and valleys (those caused by DQI) of the transmission spectra---the resonances and anti-resonances, respectively---are described by eigenvalue problems involving the molecular Hamiltonian and the self-energies.

Because DQI does not depend on the magnitude of molecule-electrode coupling, but only on where the molecule couples to the electrodes [see Eq.\ \eqref{eq:interference-eigprob}], DQI is a manifestation of substructure within the molecular Hamiltonian. The application of group theory to this substructure may reveal deeper physical and chemical insights that could lead to better molecular insulators or transistors. On a more fundamental level, such an analysis may be useful for describing chemical reactivity. It is well known in organic chemistry that \textit{meta} sites in benzene tend to be less reactive than \textit{ortho} or \textit{para} sites. Electron transport appears to follow these trends, and it would be interesting to combine this substructure analysis with transition-state theory to better understand such chemical reactions.

\begin{acknowledgments}
We thank Thorsten Hansen, Jay Bardhan, Gemma Solomon, Mark Ratner, and Justin Bergfield for helpful conversations. This work was supported by startup funds from the Institute for Advanced Computational Science at Stony Brook University.
\end{acknowledgments}

\bibliographystyle{apsrev4-1}
\bibliography{refs}

\begin{thebibliography}{48}%
\makeatletter
\providecommand \@ifxundefined [1]{%
 \@ifx{#1\undefined}
}%
\providecommand \@ifnum [1]{%
 \ifnum #1\expandafter \@firstoftwo
 \else \expandafter \@secondoftwo
 \fi
}%
\providecommand \@ifx [1]{%
 \ifx #1\expandafter \@firstoftwo
 \else \expandafter \@secondoftwo
 \fi
}%
\providecommand \natexlab [1]{#1}%
\providecommand \enquote  [1]{``#1''}%
\providecommand \bibnamefont  [1]{#1}%
\providecommand \bibfnamefont [1]{#1}%
\providecommand \citenamefont [1]{#1}%
\providecommand \href@noop [0]{\@secondoftwo}%
\providecommand \href [0]{\begingroup \@sanitize@url \@href}%
\providecommand \@href[1]{\@@startlink{#1}\@@href}%
\providecommand \@@href[1]{\endgroup#1\@@endlink}%
\providecommand \@sanitize@url [0]{\catcode `\\12\catcode `\$12\catcode
  `\&12\catcode `\#12\catcode `\^12\catcode `\_12\catcode `\%12\relax}%
\providecommand \@@startlink[1]{}%
\providecommand \@@endlink[0]{}%
\providecommand \url  [0]{\begingroup\@sanitize@url \@url }%
\providecommand \@url [1]{\endgroup\@href {#1}{\urlprefix }}%
\providecommand \urlprefix  [0]{URL }%
\providecommand \Eprint [0]{\href }%
\providecommand \doibase [0]{http://dx.doi.org/}%
\providecommand \selectlanguage [0]{\@gobble}%
\providecommand \bibinfo  [0]{\@secondoftwo}%
\providecommand \bibfield  [0]{\@secondoftwo}%
\providecommand \translation [1]{[#1]}%
\providecommand \BibitemOpen [0]{}%
\providecommand \bibitemStop [0]{}%
\providecommand \bibitemNoStop [0]{.\EOS\space}%
\providecommand \EOS [0]{\spacefactor3000\relax}%
\providecommand \BibitemShut  [1]{\csname bibitem#1\endcsname}%
\let\auto@bib@innerbib\@empty
\bibitem [{\citenamefont {Aviram}\ and\ \citenamefont
  {Ratner}(1974)}]{aviram-277-1974}%
  \BibitemOpen
  \bibfield  {author} {\bibinfo {author} {\bibfnamefont {A.}~\bibnamefont
  {Aviram}}\ and\ \bibinfo {author} {\bibfnamefont {M.~A.}\ \bibnamefont
  {Ratner}},\ }\href@noop {} {\bibfield  {journal} {\bibinfo  {journal} {Chem.
  Phys. Lett.}\ }\textbf {\bibinfo {volume} {29}},\ \bibinfo {pages} {277}
  (\bibinfo {year} {1974})}\BibitemShut {NoStop}%
\bibitem [{\citenamefont {Cuevas}\ and\ \citenamefont
  {Scheer}(2010)}]{bk:cuevas-2010}%
  \BibitemOpen
  \bibfield  {author} {\bibinfo {author} {\bibfnamefont {J.~C.}\ \bibnamefont
  {Cuevas}}\ and\ \bibinfo {author} {\bibfnamefont {E.}~\bibnamefont
  {Scheer}},\ }\href@noop {} {\emph {\bibinfo {title} {{Molecular
  Electronics}}}}\ (\bibinfo  {publisher} {World Scientific},\ \bibinfo
  {address} {Hackensack, NJ, USA},\ \bibinfo {year} {2010})\BibitemShut
  {NoStop}%
\bibitem [{\citenamefont {Bergfield}\ \emph {et~al.}(2015)\citenamefont
  {Bergfield}, \citenamefont {Heitzer}, \citenamefont {Van~Dyck}, \citenamefont
  {Marks},\ and\ \citenamefont {Ratner}}]{bergfield-6412-2015}%
  \BibitemOpen
  \bibfield  {author} {\bibinfo {author} {\bibfnamefont {J.~P.}\ \bibnamefont
  {Bergfield}}, \bibinfo {author} {\bibfnamefont {H.~M.}\ \bibnamefont
  {Heitzer}}, \bibinfo {author} {\bibfnamefont {C.}~\bibnamefont {Van~Dyck}},
  \bibinfo {author} {\bibfnamefont {T.~J.}\ \bibnamefont {Marks}}, \ and\
  \bibinfo {author} {\bibfnamefont {M.~A.}\ \bibnamefont {Ratner}},\
  }\href@noop {} {\bibfield  {journal} {\bibinfo  {journal} {ACS Nano}\
  }\textbf {\bibinfo {volume} {9}},\ \bibinfo {pages} {6412} (\bibinfo {year}
  {2015})}\BibitemShut {NoStop}%
\bibitem [{\citenamefont {Aradhya}\ \emph {et~al.}(2014)\citenamefont
  {Aradhya}, \citenamefont {Nielsen}, \citenamefont {Hybertsen},\ and\
  \citenamefont {Venkataraman}}]{aradhya-7522-2014}%
  \BibitemOpen
  \bibfield  {author} {\bibinfo {author} {\bibfnamefont {S.~V.}\ \bibnamefont
  {Aradhya}}, \bibinfo {author} {\bibfnamefont {A.}~\bibnamefont {Nielsen}},
  \bibinfo {author} {\bibfnamefont {M.~S.}\ \bibnamefont {Hybertsen}}, \ and\
  \bibinfo {author} {\bibfnamefont {L.}~\bibnamefont {Venkataraman}},\
  }\href@noop {} {\bibfield  {journal} {\bibinfo  {journal} {ACS Nano}\
  }\textbf {\bibinfo {volume} {8}},\ \bibinfo {pages} {7522} (\bibinfo {year}
  {2014})}\BibitemShut {NoStop}%
\bibitem [{\citenamefont {Li}\ \emph {et~al.}(2016)\citenamefont {Li},
  \citenamefont {Kim}, \citenamefont {Su}, \citenamefont {Steigerwald},
  \citenamefont {Nuckolls}, \citenamefont {Darancet}, \citenamefont
  {Leighton},\ and\ \citenamefont {Venkataraman}}]{li-16159-2016}%
  \BibitemOpen
  \bibfield  {author} {\bibinfo {author} {\bibfnamefont {H.}~\bibnamefont
  {Li}}, \bibinfo {author} {\bibfnamefont {N.~T.}\ \bibnamefont {Kim}},
  \bibinfo {author} {\bibfnamefont {T.~A.}\ \bibnamefont {Su}}, \bibinfo
  {author} {\bibfnamefont {M.~L.}\ \bibnamefont {Steigerwald}}, \bibinfo
  {author} {\bibfnamefont {C.}~\bibnamefont {Nuckolls}}, \bibinfo {author}
  {\bibfnamefont {P.}~\bibnamefont {Darancet}}, \bibinfo {author}
  {\bibfnamefont {J.~L.}\ \bibnamefont {Leighton}}, \ and\ \bibinfo {author}
  {\bibfnamefont {L.}~\bibnamefont {Venkataraman}},\ }\href@noop {} {\bibfield
  {journal} {\bibinfo  {journal} {J. Am. Chem. Soc.}\ }\textbf {\bibinfo
  {volume} {138}},\ \bibinfo {pages} {16159} (\bibinfo {year}
  {2016})}\BibitemShut {NoStop}%
\bibitem [{\citenamefont {Gorczak}\ \emph {et~al.}(2015)\citenamefont
  {Gorczak}, \citenamefont {Renaud}, \citenamefont {Tarkuc}, \citenamefont
  {Houtepen}, \citenamefont {Eelkema}, \citenamefont {Siebbeles},\ and\
  \citenamefont {Grozema}}]{gorczak-4196-2015}%
  \BibitemOpen
  \bibfield  {author} {\bibinfo {author} {\bibfnamefont {N.}~\bibnamefont
  {Gorczak}}, \bibinfo {author} {\bibfnamefont {N.}~\bibnamefont {Renaud}},
  \bibinfo {author} {\bibfnamefont {S.}~\bibnamefont {Tarkuc}}, \bibinfo
  {author} {\bibfnamefont {A.~J.}\ \bibnamefont {Houtepen}}, \bibinfo {author}
  {\bibfnamefont {R.}~\bibnamefont {Eelkema}}, \bibinfo {author} {\bibfnamefont
  {L.~D.~A.}\ \bibnamefont {Siebbeles}}, \ and\ \bibinfo {author}
  {\bibfnamefont {F.~C.}\ \bibnamefont {Grozema}},\ }\href@noop {} {\bibfield
  {journal} {\bibinfo  {journal} {Chem. Sci.}\ }\textbf {\bibinfo {volume}
  {6}},\ \bibinfo {pages} {4196} (\bibinfo {year} {2015})}\BibitemShut
  {NoStop}%
\bibitem [{\citenamefont {Solomon}(2015)}]{solomon-621-2015}%
  \BibitemOpen
  \bibfield  {author} {\bibinfo {author} {\bibfnamefont {G.~C.}\ \bibnamefont
  {Solomon}},\ }\href@noop {} {\bibfield  {journal} {\bibinfo  {journal} {Nat.
  Chem.}\ }\textbf {\bibinfo {volume} {7}},\ \bibinfo {pages} {621} (\bibinfo
  {year} {2015})}\BibitemShut {NoStop}%
\bibitem [{\citenamefont {Cardamone}\ \emph {et~al.}(2006)\citenamefont
  {Cardamone}, \citenamefont {Stafford},\ and\ \citenamefont
  {Mazumdar}}]{cardamone-2422-2006}%
  \BibitemOpen
  \bibfield  {author} {\bibinfo {author} {\bibfnamefont {D.~M.}\ \bibnamefont
  {Cardamone}}, \bibinfo {author} {\bibfnamefont {C.~A.}\ \bibnamefont
  {Stafford}}, \ and\ \bibinfo {author} {\bibfnamefont {S.}~\bibnamefont
  {Mazumdar}},\ }\href@noop {} {\bibfield  {journal} {\bibinfo  {journal} {Nano
  Lett.}\ }\textbf {\bibinfo {volume} {6}},\ \bibinfo {pages} {2422} (\bibinfo
  {year} {2006})}\BibitemShut {NoStop}%
\bibitem [{\citenamefont {Solomon}\ \emph
  {et~al.}(2008{\natexlab{a}})\citenamefont {Solomon}, \citenamefont {Andrews},
  \citenamefont {Hansen}, \citenamefont {Goldsmith}, \citenamefont
  {Wasielewski}, \citenamefont {Van~Duyne},\ and\ \citenamefont
  {Ratner}}]{solomon-054701-2008}%
  \BibitemOpen
  \bibfield  {author} {\bibinfo {author} {\bibfnamefont {G.~C.}\ \bibnamefont
  {Solomon}}, \bibinfo {author} {\bibfnamefont {D.~Q.}\ \bibnamefont
  {Andrews}}, \bibinfo {author} {\bibfnamefont {T.}~\bibnamefont {Hansen}},
  \bibinfo {author} {\bibfnamefont {R.~H.}\ \bibnamefont {Goldsmith}}, \bibinfo
  {author} {\bibfnamefont {M.~R.}\ \bibnamefont {Wasielewski}}, \bibinfo
  {author} {\bibfnamefont {R.~P.}\ \bibnamefont {Van~Duyne}}, \ and\ \bibinfo
  {author} {\bibfnamefont {M.~A.}\ \bibnamefont {Ratner}},\ }\href@noop {}
  {\bibfield  {journal} {\bibinfo  {journal} {J. Chem. Phys.}\ }\textbf
  {\bibinfo {volume} {129}},\ \bibinfo {pages} {054701} (\bibinfo {year}
  {2008}{\natexlab{a}})}\BibitemShut {NoStop}%
\bibitem [{\citenamefont {Hansen}\ \emph {et~al.}(2009)\citenamefont {Hansen},
  \citenamefont {Solomon}, \citenamefont {Andrews},\ and\ \citenamefont
  {Ratner}}]{hansen-194704-2009}%
  \BibitemOpen
  \bibfield  {author} {\bibinfo {author} {\bibfnamefont {T.}~\bibnamefont
  {Hansen}}, \bibinfo {author} {\bibfnamefont {G.~C.}\ \bibnamefont {Solomon}},
  \bibinfo {author} {\bibfnamefont {D.~Q.}\ \bibnamefont {Andrews}}, \ and\
  \bibinfo {author} {\bibfnamefont {M.~A.}\ \bibnamefont {Ratner}},\
  }\href@noop {} {\bibfield  {journal} {\bibinfo  {journal} {J. Chem. Phys.}\
  }\textbf {\bibinfo {volume} {131}},\ \bibinfo {pages} {194704} (\bibinfo
  {year} {2009})}\BibitemShut {NoStop}%
\bibitem [{\citenamefont {Reuter}\ and\ \citenamefont
  {Hansen}(2014)}]{reuter-181103-2014}%
  \BibitemOpen
  \bibfield  {author} {\bibinfo {author} {\bibfnamefont {M.~G.}\ \bibnamefont
  {Reuter}}\ and\ \bibinfo {author} {\bibfnamefont {T.}~\bibnamefont
  {Hansen}},\ }\href@noop {} {\bibfield  {journal} {\bibinfo  {journal} {J.
  Chem. Phys.}\ }\textbf {\bibinfo {volume} {141}},\ \bibinfo {pages} {181103}
  (\bibinfo {year} {2014})}\BibitemShut {NoStop}%
\bibitem [{\citenamefont {Kiguchi}\ \emph {et~al.}(2010)\citenamefont
  {Kiguchi}, \citenamefont {Nakamura}, \citenamefont {Takahashi}, \citenamefont
  {Takahashi},\ and\ \citenamefont {Ohto}}]{kiguchi-22254-2010}%
  \BibitemOpen
  \bibfield  {author} {\bibinfo {author} {\bibfnamefont {M.}~\bibnamefont
  {Kiguchi}}, \bibinfo {author} {\bibfnamefont {H.}~\bibnamefont {Nakamura}},
  \bibinfo {author} {\bibfnamefont {Y.}~\bibnamefont {Takahashi}}, \bibinfo
  {author} {\bibfnamefont {T.}~\bibnamefont {Takahashi}}, \ and\ \bibinfo
  {author} {\bibfnamefont {T.}~\bibnamefont {Ohto}},\ }\href@noop {} {\bibfield
   {journal} {\bibinfo  {journal} {J. Phys. Chem. C}\ }\textbf {\bibinfo
  {volume} {114}},\ \bibinfo {pages} {22254} (\bibinfo {year}
  {2010})}\BibitemShut {NoStop}%
\bibitem [{\citenamefont {Fracasso}\ \emph {et~al.}(2011)\citenamefont
  {Fracasso}, \citenamefont {Valkenier}, \citenamefont {Hummelen},
  \citenamefont {Solomon},\ and\ \citenamefont {Chiechi}}]{fracasso-9556-2011}%
  \BibitemOpen
  \bibfield  {author} {\bibinfo {author} {\bibfnamefont {D.}~\bibnamefont
  {Fracasso}}, \bibinfo {author} {\bibfnamefont {H.}~\bibnamefont {Valkenier}},
  \bibinfo {author} {\bibfnamefont {J.~C.}\ \bibnamefont {Hummelen}}, \bibinfo
  {author} {\bibfnamefont {G.~C.}\ \bibnamefont {Solomon}}, \ and\ \bibinfo
  {author} {\bibfnamefont {R.~C.}\ \bibnamefont {Chiechi}},\ }\href@noop {}
  {\bibfield  {journal} {\bibinfo  {journal} {J. Am. Chem. Soc.}\ }\textbf
  {\bibinfo {volume} {133}},\ \bibinfo {pages} {9556} (\bibinfo {year}
  {2011})}\BibitemShut {NoStop}%
\bibitem [{\citenamefont {Taniguchi}\ \emph {et~al.}(2011)\citenamefont
  {Taniguchi}, \citenamefont {Tsutsui}, \citenamefont {Mogi}, \citenamefont
  {Sugawara}, \citenamefont {Tsuji}, \citenamefont {Yoshizawa},\ and\
  \citenamefont {Kawai}}]{taniguchi-11426-2011}%
  \BibitemOpen
  \bibfield  {author} {\bibinfo {author} {\bibfnamefont {M.}~\bibnamefont
  {Taniguchi}}, \bibinfo {author} {\bibfnamefont {M.}~\bibnamefont {Tsutsui}},
  \bibinfo {author} {\bibfnamefont {R.}~\bibnamefont {Mogi}}, \bibinfo {author}
  {\bibfnamefont {T.}~\bibnamefont {Sugawara}}, \bibinfo {author}
  {\bibfnamefont {Y.}~\bibnamefont {Tsuji}}, \bibinfo {author} {\bibfnamefont
  {K.}~\bibnamefont {Yoshizawa}}, \ and\ \bibinfo {author} {\bibfnamefont
  {T.}~\bibnamefont {Kawai}},\ }\href@noop {} {\bibfield  {journal} {\bibinfo
  {journal} {J. Am. Chem. Soc.}\ }\textbf {\bibinfo {volume} {133}},\ \bibinfo
  {pages} {11426} (\bibinfo {year} {2011})}\BibitemShut {NoStop}%
\bibitem [{\citenamefont {Gu{\'e}don}\ \emph {et~al.}(2012)\citenamefont
  {Gu{\'e}don}, \citenamefont {Valkenier}, \citenamefont {Markussen},
  \citenamefont {Thygesen}, \citenamefont {Hummelen},\ and\ \citenamefont
  {van~der Molen}}]{guedon-305-2012}%
  \BibitemOpen
  \bibfield  {author} {\bibinfo {author} {\bibfnamefont {C.~M.}\ \bibnamefont
  {Gu{\'e}don}}, \bibinfo {author} {\bibfnamefont {H.}~\bibnamefont
  {Valkenier}}, \bibinfo {author} {\bibfnamefont {T.}~\bibnamefont
  {Markussen}}, \bibinfo {author} {\bibfnamefont {K.~S.}\ \bibnamefont
  {Thygesen}}, \bibinfo {author} {\bibfnamefont {J.~C.}\ \bibnamefont
  {Hummelen}}, \ and\ \bibinfo {author} {\bibfnamefont {S.~J.}\ \bibnamefont
  {van~der Molen}},\ }\href@noop {} {\bibfield  {journal} {\bibinfo  {journal}
  {Nat. Nanotech.}\ }\textbf {\bibinfo {volume} {7}},\ \bibinfo {pages} {305}
  (\bibinfo {year} {2012})}\BibitemShut {NoStop}%
\bibitem [{\citenamefont {Aradhya}\ \emph {et~al.}(2012)\citenamefont
  {Aradhya}, \citenamefont {Meisner}, \citenamefont {Krikorian}, \citenamefont
  {Ahn}, \citenamefont {Parameswaran}, \citenamefont {Steigerwald},
  \citenamefont {Nuckolls},\ and\ \citenamefont
  {Venkataraman}}]{aradhya-1643-2012}%
  \BibitemOpen
  \bibfield  {author} {\bibinfo {author} {\bibfnamefont {S.~V.}\ \bibnamefont
  {Aradhya}}, \bibinfo {author} {\bibfnamefont {J.~S.}\ \bibnamefont
  {Meisner}}, \bibinfo {author} {\bibfnamefont {M.}~\bibnamefont {Krikorian}},
  \bibinfo {author} {\bibfnamefont {S.}~\bibnamefont {Ahn}}, \bibinfo {author}
  {\bibfnamefont {R.}~\bibnamefont {Parameswaran}}, \bibinfo {author}
  {\bibfnamefont {M.~L.}\ \bibnamefont {Steigerwald}}, \bibinfo {author}
  {\bibfnamefont {C.}~\bibnamefont {Nuckolls}}, \ and\ \bibinfo {author}
  {\bibfnamefont {L.}~\bibnamefont {Venkataraman}},\ }\href@noop {} {\bibfield
  {journal} {\bibinfo  {journal} {Nano Lett.}\ }\textbf {\bibinfo {volume}
  {12}},\ \bibinfo {pages} {1643} (\bibinfo {year} {2012})}\BibitemShut
  {NoStop}%
\bibitem [{\citenamefont {Arroyo}\ \emph {et~al.}(2013)\citenamefont {Arroyo},
  \citenamefont {Tarkuc}, \citenamefont {Frisenda}, \citenamefont
  {Seldenthuis}, \citenamefont {Woerde}, \citenamefont {Eelkema}, \citenamefont
  {Grozema},\ and\ \citenamefont {van~der Zant}}]{arroyo-3152-2013}%
  \BibitemOpen
  \bibfield  {author} {\bibinfo {author} {\bibfnamefont {C.~R.}\ \bibnamefont
  {Arroyo}}, \bibinfo {author} {\bibfnamefont {S.}~\bibnamefont {Tarkuc}},
  \bibinfo {author} {\bibfnamefont {R.}~\bibnamefont {Frisenda}}, \bibinfo
  {author} {\bibfnamefont {J.~S.}\ \bibnamefont {Seldenthuis}}, \bibinfo
  {author} {\bibfnamefont {C.~H.~M.}\ \bibnamefont {Woerde}}, \bibinfo {author}
  {\bibfnamefont {R.}~\bibnamefont {Eelkema}}, \bibinfo {author} {\bibfnamefont
  {F.~C.}\ \bibnamefont {Grozema}}, \ and\ \bibinfo {author} {\bibfnamefont
  {H.~S.~J.}\ \bibnamefont {van~der Zant}},\ }\href@noop {} {\bibfield
  {journal} {\bibinfo  {journal} {Angew. Chem. Int. Ed.}\ }\textbf {\bibinfo
  {volume} {52}},\ \bibinfo {pages} {3152} (\bibinfo {year}
  {2013})}\BibitemShut {NoStop}%
\bibitem [{\citenamefont {Liu}\ \emph {et~al.}(2017)\citenamefont {Liu},
  \citenamefont {Sangtarash}, \citenamefont {Reber}, \citenamefont {Zhang},
  \citenamefont {Sadeghi}, \citenamefont {Shi}, \citenamefont {Xiao},
  \citenamefont {Hong}, \citenamefont {Lambert},\ and\ \citenamefont
  {Liu}}]{liu-173-2017}%
  \BibitemOpen
  \bibfield  {author} {\bibinfo {author} {\bibfnamefont {X.}~\bibnamefont
  {Liu}}, \bibinfo {author} {\bibfnamefont {S.}~\bibnamefont {Sangtarash}},
  \bibinfo {author} {\bibfnamefont {D.}~\bibnamefont {Reber}}, \bibinfo
  {author} {\bibfnamefont {D.}~\bibnamefont {Zhang}}, \bibinfo {author}
  {\bibfnamefont {H.}~\bibnamefont {Sadeghi}}, \bibinfo {author} {\bibfnamefont
  {J.}~\bibnamefont {Shi}}, \bibinfo {author} {\bibfnamefont {Z.-Y.}\
  \bibnamefont {Xiao}}, \bibinfo {author} {\bibfnamefont {W.}~\bibnamefont
  {Hong}}, \bibinfo {author} {\bibfnamefont {C.~J.}\ \bibnamefont {Lambert}}, \
  and\ \bibinfo {author} {\bibfnamefont {S.-X.}\ \bibnamefont {Liu}},\
  }\href@noop {} {\bibfield  {journal} {\bibinfo  {journal} {Angew. Chem. Int.
  Ed.}\ }\textbf {\bibinfo {volume} {56}},\ \bibinfo {pages} {173} (\bibinfo
  {year} {2017})}\BibitemShut {NoStop}%
\bibitem [{\citenamefont {Morikawa}\ \emph {et~al.}(2005)\citenamefont
  {Morikawa}, \citenamefont {Narita},\ and\ \citenamefont
  {Klein}}]{morikawa-554-2005}%
  \BibitemOpen
  \bibfield  {author} {\bibinfo {author} {\bibfnamefont {T.}~\bibnamefont
  {Morikawa}}, \bibinfo {author} {\bibfnamefont {S.}~\bibnamefont {Narita}}, \
  and\ \bibinfo {author} {\bibfnamefont {D.~J.}\ \bibnamefont {Klein}},\
  }\href@noop {} {\bibfield  {journal} {\bibinfo  {journal} {Chem. Phys.
  Lett.}\ }\textbf {\bibinfo {volume} {402}},\ \bibinfo {pages} {554} (\bibinfo
  {year} {2005})}\BibitemShut {NoStop}%
\bibitem [{\citenamefont {Pickup}\ and\ \citenamefont
  {Fowler}(2008)}]{pickup-198-2008}%
  \BibitemOpen
  \bibfield  {author} {\bibinfo {author} {\bibfnamefont {B.~T.}\ \bibnamefont
  {Pickup}}\ and\ \bibinfo {author} {\bibfnamefont {P.~W.}\ \bibnamefont
  {Fowler}},\ }\href@noop {} {\bibfield  {journal} {\bibinfo  {journal} {Chem.
  Phys. Lett.}\ }\textbf {\bibinfo {volume} {459}},\ \bibinfo {pages} {198}
  (\bibinfo {year} {2008})}\BibitemShut {NoStop}%
\bibitem [{\citenamefont {Yoshizawa}\ \emph {et~al.}(2008)\citenamefont
  {Yoshizawa}, \citenamefont {Tada},\ and\ \citenamefont
  {Staykov}}]{yoshizawa-9406-2008}%
  \BibitemOpen
  \bibfield  {author} {\bibinfo {author} {\bibfnamefont {K.}~\bibnamefont
  {Yoshizawa}}, \bibinfo {author} {\bibfnamefont {T.}~\bibnamefont {Tada}}, \
  and\ \bibinfo {author} {\bibfnamefont {A.}~\bibnamefont {Staykov}},\
  }\href@noop {} {\bibfield  {journal} {\bibinfo  {journal} {J. Am. Chem.
  Soc.}\ }\textbf {\bibinfo {volume} {130}},\ \bibinfo {pages} {9406} (\bibinfo
  {year} {2008})}\BibitemShut {NoStop}%
\bibitem [{\citenamefont {Fowler}\ \emph
  {et~al.}(2009{\natexlab{a}})\citenamefont {Fowler}, \citenamefont {Pickup},
  \citenamefont {Todorova},\ and\ \citenamefont
  {Myrvold}}]{fowler-044104-2009}%
  \BibitemOpen
  \bibfield  {author} {\bibinfo {author} {\bibfnamefont {P.~W.}\ \bibnamefont
  {Fowler}}, \bibinfo {author} {\bibfnamefont {B.~T.}\ \bibnamefont {Pickup}},
  \bibinfo {author} {\bibfnamefont {T.~Z.}\ \bibnamefont {Todorova}}, \ and\
  \bibinfo {author} {\bibfnamefont {W.}~\bibnamefont {Myrvold}},\ }\href@noop
  {} {\bibfield  {journal} {\bibinfo  {journal} {J. Chem. Phys.}\ }\textbf
  {\bibinfo {volume} {131}},\ \bibinfo {pages} {044104} (\bibinfo {year}
  {2009}{\natexlab{a}})}\BibitemShut {NoStop}%
\bibitem [{\citenamefont {Fowler}\ \emph
  {et~al.}(2009{\natexlab{b}})\citenamefont {Fowler}, \citenamefont {Pickup},
  \citenamefont {Todorova},\ and\ \citenamefont
  {Myrvold}}]{fowler-244110-2009}%
  \BibitemOpen
  \bibfield  {author} {\bibinfo {author} {\bibfnamefont {P.~W.}\ \bibnamefont
  {Fowler}}, \bibinfo {author} {\bibfnamefont {B.~T.}\ \bibnamefont {Pickup}},
  \bibinfo {author} {\bibfnamefont {T.~Z.}\ \bibnamefont {Todorova}}, \ and\
  \bibinfo {author} {\bibfnamefont {W.}~\bibnamefont {Myrvold}},\ }\href@noop
  {} {\bibfield  {journal} {\bibinfo  {journal} {J. Chem. Phys.}\ }\textbf
  {\bibinfo {volume} {131}},\ \bibinfo {pages} {244110} (\bibinfo {year}
  {2009}{\natexlab{b}})}\BibitemShut {NoStop}%
\bibitem [{\citenamefont {Yoshizawa}(2012)}]{yoshizawa-1612-2012}%
  \BibitemOpen
  \bibfield  {author} {\bibinfo {author} {\bibfnamefont {K.}~\bibnamefont
  {Yoshizawa}},\ }\href@noop {} {\bibfield  {journal} {\bibinfo  {journal}
  {Acc. Chem. Res.}\ }\textbf {\bibinfo {volume} {45}},\ \bibinfo {pages}
  {1612} (\bibinfo {year} {2012})}\BibitemShut {NoStop}%
\bibitem [{\citenamefont {Tsuji}\ \emph {et~al.}(2014)\citenamefont {Tsuji},
  \citenamefont {Hoffmann}, \citenamefont {Movassagh},\ and\ \citenamefont
  {Datta}}]{tsuji-224311-2014}%
  \BibitemOpen
  \bibfield  {author} {\bibinfo {author} {\bibfnamefont {Y.}~\bibnamefont
  {Tsuji}}, \bibinfo {author} {\bibfnamefont {R.}~\bibnamefont {Hoffmann}},
  \bibinfo {author} {\bibfnamefont {R.}~\bibnamefont {Movassagh}}, \ and\
  \bibinfo {author} {\bibfnamefont {S.}~\bibnamefont {Datta}},\ }\href@noop {}
  {\bibfield  {journal} {\bibinfo  {journal} {J. Chem. Phys.}\ }\textbf
  {\bibinfo {volume} {141}},\ \bibinfo {pages} {224311} (\bibinfo {year}
  {2014})}\BibitemShut {NoStop}%
\bibitem [{\citenamefont {Stuyver}\ \emph {et~al.}(2015)\citenamefont
  {Stuyver}, \citenamefont {Fias}, \citenamefont {De~Proft},\ and\
  \citenamefont {Geerlings}}]{stuyver-26390-2015}%
  \BibitemOpen
  \bibfield  {author} {\bibinfo {author} {\bibfnamefont {T.}~\bibnamefont
  {Stuyver}}, \bibinfo {author} {\bibfnamefont {S.}~\bibnamefont {Fias}},
  \bibinfo {author} {\bibfnamefont {F.}~\bibnamefont {De~Proft}}, \ and\
  \bibinfo {author} {\bibfnamefont {P.}~\bibnamefont {Geerlings}},\ }\href@noop
  {} {\bibfield  {journal} {\bibinfo  {journal} {J. Phys. Chem. C}\ }\textbf
  {\bibinfo {volume} {119}},\ \bibinfo {pages} {26390} (\bibinfo {year}
  {2015})}\BibitemShut {NoStop}%
\bibitem [{\citenamefont {Markussen}\ \emph {et~al.}(2010)\citenamefont
  {Markussen}, \citenamefont {Stadler},\ and\ \citenamefont
  {Thygesen}}]{markussen-4260-2010}%
  \BibitemOpen
  \bibfield  {author} {\bibinfo {author} {\bibfnamefont {T.}~\bibnamefont
  {Markussen}}, \bibinfo {author} {\bibfnamefont {R.}~\bibnamefont {Stadler}},
  \ and\ \bibinfo {author} {\bibfnamefont {K.~S.}\ \bibnamefont {Thygesen}},\
  }\href@noop {} {\bibfield  {journal} {\bibinfo  {journal} {Nano Lett.}\
  }\textbf {\bibinfo {volume} {10}},\ \bibinfo {pages} {4260} (\bibinfo {year}
  {2010})}\BibitemShut {NoStop}%
\bibitem [{\citenamefont {Markussen}\ \emph {et~al.}(2011)\citenamefont
  {Markussen}, \citenamefont {Stadler},\ and\ \citenamefont
  {Thygesen}}]{markussen-14311-2011}%
  \BibitemOpen
  \bibfield  {author} {\bibinfo {author} {\bibfnamefont {T.}~\bibnamefont
  {Markussen}}, \bibinfo {author} {\bibfnamefont {R.}~\bibnamefont {Stadler}},
  \ and\ \bibinfo {author} {\bibfnamefont {K.~S.}\ \bibnamefont {Thygesen}},\
  }\href@noop {} {\bibfield  {journal} {\bibinfo  {journal} {Phys. Chem. Chem.
  Phys.}\ }\textbf {\bibinfo {volume} {13}},\ \bibinfo {pages} {14311}
  (\bibinfo {year} {2011})}\BibitemShut {NoStop}%
\bibitem [{\citenamefont {Nozaki}\ \emph {et~al.}(2013)\citenamefont {Nozaki},
  \citenamefont {Sevin{\c c}li}, \citenamefont {Avdoshenko}, \citenamefont
  {Gutierrez},\ and\ \citenamefont {Cuniberti}}]{nozaki-13951-2013}%
  \BibitemOpen
  \bibfield  {author} {\bibinfo {author} {\bibfnamefont {D.}~\bibnamefont
  {Nozaki}}, \bibinfo {author} {\bibfnamefont {H.}~\bibnamefont {Sevin{\c
  c}li}}, \bibinfo {author} {\bibfnamefont {S.~M.}\ \bibnamefont {Avdoshenko}},
  \bibinfo {author} {\bibfnamefont {R.}~\bibnamefont {Gutierrez}}, \ and\
  \bibinfo {author} {\bibfnamefont {G.}~\bibnamefont {Cuniberti}},\ }\href@noop
  {} {\bibfield  {journal} {\bibinfo  {journal} {Phys. Chem. Chem. Phys.}\
  }\textbf {\bibinfo {volume} {15}},\ \bibinfo {pages} {13951} (\bibinfo {year}
  {2013})}\BibitemShut {NoStop}%
\bibitem [{\citenamefont {Zhao}\ \emph {et~al.}(2016)\citenamefont {Zhao},
  \citenamefont {Geskin},\ and\ \citenamefont {Stadler}}]{zhao-092308-2016}%
  \BibitemOpen
  \bibfield  {author} {\bibinfo {author} {\bibfnamefont {X.}~\bibnamefont
  {Zhao}}, \bibinfo {author} {\bibfnamefont {V.}~\bibnamefont {Geskin}}, \ and\
  \bibinfo {author} {\bibfnamefont {R.}~\bibnamefont {Stadler}},\ }\href@noop
  {} {\bibfield  {journal} {\bibinfo  {journal} {J. Chem. Phys.}\ }\textbf
  {\bibinfo {volume} {146}},\ \bibinfo {pages} {092308} (\bibinfo {year}
  {2016})}\BibitemShut {NoStop}%
\bibitem [{\citenamefont {Xia}\ \emph {et~al.}(2014)\citenamefont {Xia},
  \citenamefont {Capozzi}, \citenamefont {Wei}, \citenamefont {Strange},
  \citenamefont {Batra}, \citenamefont {Moreno}, \citenamefont {Amir},
  \citenamefont {Amir}, \citenamefont {Solomon}, \citenamefont {Venkataraman},\
  and\ \citenamefont {Campos}}]{xia-2941-2014}%
  \BibitemOpen
  \bibfield  {author} {\bibinfo {author} {\bibfnamefont {J.}~\bibnamefont
  {Xia}}, \bibinfo {author} {\bibfnamefont {B.}~\bibnamefont {Capozzi}},
  \bibinfo {author} {\bibfnamefont {S.}~\bibnamefont {Wei}}, \bibinfo {author}
  {\bibfnamefont {M.}~\bibnamefont {Strange}}, \bibinfo {author} {\bibfnamefont
  {A.}~\bibnamefont {Batra}}, \bibinfo {author} {\bibfnamefont {J.~R.}\
  \bibnamefont {Moreno}}, \bibinfo {author} {\bibfnamefont {R.~J.}\
  \bibnamefont {Amir}}, \bibinfo {author} {\bibfnamefont {E.}~\bibnamefont
  {Amir}}, \bibinfo {author} {\bibfnamefont {G.~C.}\ \bibnamefont {Solomon}},
  \bibinfo {author} {\bibfnamefont {L.}~\bibnamefont {Venkataraman}}, \ and\
  \bibinfo {author} {\bibfnamefont {L.~M.}\ \bibnamefont {Campos}},\
  }\href@noop {} {\bibfield  {journal} {\bibinfo  {journal} {Nano Lett.}\
  }\textbf {\bibinfo {volume} {14}},\ \bibinfo {pages} {2941} (\bibinfo {year}
  {2014})}\BibitemShut {NoStop}%
\bibitem [{\citenamefont {Bergfield}\ \emph {et~al.}(2010)\citenamefont
  {Bergfield}, \citenamefont {Solis},\ and\ \citenamefont
  {Stafford}}]{bergfield-5314-2010}%
  \BibitemOpen
  \bibfield  {author} {\bibinfo {author} {\bibfnamefont {J.~P.}\ \bibnamefont
  {Bergfield}}, \bibinfo {author} {\bibfnamefont {M.~A.}\ \bibnamefont
  {Solis}}, \ and\ \bibinfo {author} {\bibfnamefont {C.~A.}\ \bibnamefont
  {Stafford}},\ }\href@noop {} {\bibfield  {journal} {\bibinfo  {journal} {ACS
  Nano}\ }\textbf {\bibinfo {volume} {4}},\ \bibinfo {pages} {5314} (\bibinfo
  {year} {2010})}\BibitemShut {NoStop}%
\bibitem [{\citenamefont {Hansen}\ and\ \citenamefont
  {Solomon}(2016)}]{hansen-6295-2016}%
  \BibitemOpen
  \bibfield  {author} {\bibinfo {author} {\bibfnamefont {T.}~\bibnamefont
  {Hansen}}\ and\ \bibinfo {author} {\bibfnamefont {G.~C.}\ \bibnamefont
  {Solomon}},\ }\href@noop {} {\bibfield  {journal} {\bibinfo  {journal} {J.
  Phys. Chem. C}\ }\textbf {\bibinfo {volume} {120}},\ \bibinfo {pages} {6295}
  (\bibinfo {year} {2016})}\BibitemShut {NoStop}%
\bibitem [{\citenamefont {B{\"u}ttiker}\ \emph {et~al.}(1985)\citenamefont
  {B{\"u}ttiker}, \citenamefont {Imry}, \citenamefont {Landauer},\ and\
  \citenamefont {Pinhas}}]{buttiker-6207-1985}%
  \BibitemOpen
  \bibfield  {author} {\bibinfo {author} {\bibfnamefont {M.}~\bibnamefont
  {B{\"u}ttiker}}, \bibinfo {author} {\bibfnamefont {Y.}~\bibnamefont {Imry}},
  \bibinfo {author} {\bibfnamefont {R.}~\bibnamefont {Landauer}}, \ and\
  \bibinfo {author} {\bibfnamefont {S.}~\bibnamefont {Pinhas}},\ }\href@noop {}
  {\bibfield  {journal} {\bibinfo  {journal} {Phys. Rev. B}\ }\textbf {\bibinfo
  {volume} {31}},\ \bibinfo {pages} {6207} (\bibinfo {year}
  {1985})}\BibitemShut {NoStop}%
\bibitem [{\citenamefont {Imry}\ and\ \citenamefont
  {Landauer}(1999)}]{imry-s306-1999}%
  \BibitemOpen
  \bibfield  {author} {\bibinfo {author} {\bibfnamefont {Y.}~\bibnamefont
  {Imry}}\ and\ \bibinfo {author} {\bibfnamefont {R.}~\bibnamefont
  {Landauer}},\ }\href@noop {} {\bibfield  {journal} {\bibinfo  {journal} {Rev.
  Mod. Phys.}\ }\textbf {\bibinfo {volume} {71}},\ \bibinfo {pages} {S306}
  (\bibinfo {year} {1999})}\BibitemShut {NoStop}%
\bibitem [{\citenamefont {Taylor}\ \emph {et~al.}(2001)\citenamefont {Taylor},
  \citenamefont {Guo},\ and\ \citenamefont {Wang}}]{taylor-245407-2001}%
  \BibitemOpen
  \bibfield  {author} {\bibinfo {author} {\bibfnamefont {J.}~\bibnamefont
  {Taylor}}, \bibinfo {author} {\bibfnamefont {H.}~\bibnamefont {Guo}}, \ and\
  \bibinfo {author} {\bibfnamefont {J.}~\bibnamefont {Wang}},\ }\href@noop {}
  {\bibfield  {journal} {\bibinfo  {journal} {Phys. Rev. B}\ }\textbf {\bibinfo
  {volume} {63}},\ \bibinfo {pages} {245407} (\bibinfo {year}
  {2001})}\BibitemShut {NoStop}%
\bibitem [{\citenamefont {Dhar}\ and\ \citenamefont
  {Sen}(2006)}]{dhar-085119-2006}%
  \BibitemOpen
  \bibfield  {author} {\bibinfo {author} {\bibfnamefont {A.}~\bibnamefont
  {Dhar}}\ and\ \bibinfo {author} {\bibfnamefont {D.}~\bibnamefont {Sen}},\
  }\href@noop {} {\bibfield  {journal} {\bibinfo  {journal} {Phys. Rev. B}\
  }\textbf {\bibinfo {volume} {73}},\ \bibinfo {pages} {085119} (\bibinfo
  {year} {2006})}\BibitemShut {NoStop}%
\bibitem [{\citenamefont {Bowen}\ \emph {et~al.}(1995)\citenamefont {Bowen},
  \citenamefont {Frensley}, \citenamefont {Klimeck},\ and\ \citenamefont
  {Lake}}]{bowen-2754-1995}%
  \BibitemOpen
  \bibfield  {author} {\bibinfo {author} {\bibfnamefont {R.~C.}\ \bibnamefont
  {Bowen}}, \bibinfo {author} {\bibfnamefont {W.~R.}\ \bibnamefont {Frensley}},
  \bibinfo {author} {\bibfnamefont {G.}~\bibnamefont {Klimeck}}, \ and\
  \bibinfo {author} {\bibfnamefont {R.~K.}\ \bibnamefont {Lake}},\ }\href@noop
  {} {\bibfield  {journal} {\bibinfo  {journal} {Phys. Rev. B}\ }\textbf
  {\bibinfo {volume} {52}},\ \bibinfo {pages} {2754} (\bibinfo {year}
  {1995})}\BibitemShut {NoStop}%
\bibitem [{\citenamefont {Solomon}\ \emph
  {et~al.}(2008{\natexlab{b}})\citenamefont {Solomon}, \citenamefont {Andrews},
  \citenamefont {Goldsmith}, \citenamefont {Hansen}, \citenamefont
  {Wasielewski}, \citenamefont {Van~Duyne},\ and\ \citenamefont
  {Ratner}}]{solomon-17301-2008}%
  \BibitemOpen
  \bibfield  {author} {\bibinfo {author} {\bibfnamefont {G.~C.}\ \bibnamefont
  {Solomon}}, \bibinfo {author} {\bibfnamefont {D.~Q.}\ \bibnamefont
  {Andrews}}, \bibinfo {author} {\bibfnamefont {R.~H.}\ \bibnamefont
  {Goldsmith}}, \bibinfo {author} {\bibfnamefont {T.}~\bibnamefont {Hansen}},
  \bibinfo {author} {\bibfnamefont {M.~R.}\ \bibnamefont {Wasielewski}},
  \bibinfo {author} {\bibfnamefont {R.~P.}\ \bibnamefont {Van~Duyne}}, \ and\
  \bibinfo {author} {\bibfnamefont {M.~A.}\ \bibnamefont {Ratner}},\
  }\href@noop {} {\bibfield  {journal} {\bibinfo  {journal} {J. Am. Chem.
  Soc.}\ }\textbf {\bibinfo {volume} {130}},\ \bibinfo {pages} {17301}
  (\bibinfo {year} {2008}{\natexlab{b}})}\BibitemShut {NoStop}%
\bibitem [{Note1()}]{Note1}%
  \BibitemOpen
  \bibinfo {note} {Note that $\protect \mathbf {\Gamma }_\protect \mathrm
  {L/R}(E)$ is essentially the anti-Hermitian part of $\protect \mathbf {\Sigma
  }_\protect \mathrm {L/R}(E)$, such that only the Hermitian parts of $\protect
  \mathbf {\Sigma }_\protect \mathrm {L}(E)$ and $\protect \mathbf {\Sigma
  }_\protect \mathrm {R}(E)$ survive the restriction. Because the Hermitian and
  anti-Hermitian parts of $\protect \mathbf {\Sigma }_\protect \mathrm
  {L/R}(E)$ are related by a Kramers-Kronig relation (\protect \textit {i.e.},\
  Hilbert transform) \cite {bk:economou-2006}, the kernel of the Hermitian part
  will contain the kernel of the anti-Hermitian part except at rare and
  isolated energies. We also assume that $\protect \text {Ker}[\protect \mathbf
  {\Gamma }_\protect \mathrm {L/R}(E)]$ does not depend on $E$, meaning that
  the parts of the molecule coupled to the electrodes do not change with
  $E$.}\BibitemShut {Stop}%
\bibitem [{\citenamefont {Wilkinson}(1979)}]{wilkinson-285-1979}%
  \BibitemOpen
  \bibfield  {author} {\bibinfo {author} {\bibfnamefont {J.~H.}\ \bibnamefont
  {Wilkinson}},\ }\href@noop {} {\bibfield  {journal} {\bibinfo  {journal}
  {Lin. Alg. Appl.}\ }\textbf {\bibinfo {volume} {28}},\ \bibinfo {pages} {285}
  (\bibinfo {year} {1979})}\BibitemShut {NoStop}%
\bibitem [{\citenamefont {Van~Dooren}(1979)}]{van-dooren-103-1979}%
  \BibitemOpen
  \bibfield  {author} {\bibinfo {author} {\bibfnamefont {P.}~\bibnamefont
  {Van~Dooren}},\ }\href@noop {} {\bibfield  {journal} {\bibinfo  {journal}
  {Lin. Alg. Appl.}\ }\textbf {\bibinfo {volume} {27}},\ \bibinfo {pages} {103}
  (\bibinfo {year} {1979})}\BibitemShut {NoStop}%
\bibitem [{\citenamefont {Gustafson}(2012)}]{bk:gustafson-2012}%
  \BibitemOpen
  \bibfield  {author} {\bibinfo {author} {\bibfnamefont {K.}~\bibnamefont
  {Gustafson}},\ }\href@noop {} {\emph {\bibinfo {title} {{Antieigenvalue
  Analysis}}}}\ (\bibinfo  {publisher} {World Scientific},\ \bibinfo {address}
  {Hackensack, NJ, USA},\ \bibinfo {year} {2012})\BibitemShut {NoStop}%
\bibitem [{\citenamefont {Verzijl}\ \emph {et~al.}(2013)\citenamefont
  {Verzijl}, \citenamefont {Seldenthuis},\ and\ \citenamefont
  {Thijssen}}]{verzijl-094102-2013}%
  \BibitemOpen
  \bibfield  {author} {\bibinfo {author} {\bibfnamefont {C.~J.~O.}\
  \bibnamefont {Verzijl}}, \bibinfo {author} {\bibfnamefont {J.~S.}\
  \bibnamefont {Seldenthuis}}, \ and\ \bibinfo {author} {\bibfnamefont {J.~M.}\
  \bibnamefont {Thijssen}},\ }\href@noop {} {\bibfield  {journal} {\bibinfo
  {journal} {J. Chem. Phys.}\ }\textbf {\bibinfo {volume} {138}},\ \bibinfo
  {pages} {094102} (\bibinfo {year} {2013})}\BibitemShut {NoStop}%
\bibitem [{\citenamefont {Fowler}\ \emph
  {et~al.}(2009{\natexlab{c}})\citenamefont {Fowler}, \citenamefont {Pickup},
  \citenamefont {Todorova},\ and\ \citenamefont
  {Pisanski}}]{fowler-174708-2009}%
  \BibitemOpen
  \bibfield  {author} {\bibinfo {author} {\bibfnamefont {P.~W.}\ \bibnamefont
  {Fowler}}, \bibinfo {author} {\bibfnamefont {B.~T.}\ \bibnamefont {Pickup}},
  \bibinfo {author} {\bibfnamefont {T.~Z.}\ \bibnamefont {Todorova}}, \ and\
  \bibinfo {author} {\bibfnamefont {T.}~\bibnamefont {Pisanski}},\ }\href@noop
  {} {\bibfield  {journal} {\bibinfo  {journal} {J. Chem. Phys.}\ }\textbf
  {\bibinfo {volume} {130}},\ \bibinfo {pages} {174708} (\bibinfo {year}
  {2009}{\natexlab{c}})}\BibitemShut {NoStop}%
\bibitem [{\citenamefont {Andrews}\ \emph {et~al.}(2008)\citenamefont
  {Andrews}, \citenamefont {Solomon}, \citenamefont {Van~Duyne},\ and\
  \citenamefont {Ratner}}]{andrews-17309-2008}%
  \BibitemOpen
  \bibfield  {author} {\bibinfo {author} {\bibfnamefont {D.~Q.}\ \bibnamefont
  {Andrews}}, \bibinfo {author} {\bibfnamefont {G.~C.}\ \bibnamefont
  {Solomon}}, \bibinfo {author} {\bibfnamefont {R.~P.}\ \bibnamefont
  {Van~Duyne}}, \ and\ \bibinfo {author} {\bibfnamefont {M.~A.}\ \bibnamefont
  {Ratner}},\ }\href@noop {} {\bibfield  {journal} {\bibinfo  {journal} {J. Am.
  Chem. Soc.}\ }\textbf {\bibinfo {volume} {130}},\ \bibinfo {pages} {17309}
  (\bibinfo {year} {2008})}\BibitemShut {NoStop}%
\bibitem [{\citenamefont {Pedersen}\ \emph {et~al.}(2015)\citenamefont
  {Pedersen}, \citenamefont {Borges}, \citenamefont {Hedeg{\aa}rd},
  \citenamefont {Solomon},\ and\ \citenamefont
  {Strange}}]{pedersen-26919-2015}%
  \BibitemOpen
  \bibfield  {author} {\bibinfo {author} {\bibfnamefont {K.~G.~L.}\
  \bibnamefont {Pedersen}}, \bibinfo {author} {\bibfnamefont {A.}~\bibnamefont
  {Borges}}, \bibinfo {author} {\bibfnamefont {P.}~\bibnamefont
  {Hedeg{\aa}rd}}, \bibinfo {author} {\bibfnamefont {G.~C.}\ \bibnamefont
  {Solomon}}, \ and\ \bibinfo {author} {\bibfnamefont {M.}~\bibnamefont
  {Strange}},\ }\href@noop {} {\bibfield  {journal} {\bibinfo  {journal} {J.
  Phys. Chem. C}\ }\textbf {\bibinfo {volume} {119}},\ \bibinfo {pages} {26919}
  (\bibinfo {year} {2015})}\BibitemShut {NoStop}%
\bibitem [{\citenamefont {Economou}(2006)}]{bk:economou-2006}%
  \BibitemOpen
  \bibfield  {author} {\bibinfo {author} {\bibfnamefont {E.~N.}\ \bibnamefont
  {Economou}},\ }\href@noop {} {\emph {\bibinfo {title} {{Green's Functions in
  Quantum Physics}}}},\ \bibinfo {edition} {3rd}\ ed.\ (\bibinfo  {publisher}
  {Springer-Verlag},\ \bibinfo {address} {Heidelberg, Germany},\ \bibinfo
  {year} {2006})\BibitemShut {NoStop}%
\end{thebibliography}%

\end{document}